\documentclass{aa}
\usepackage{psfig,latexsym,longtable,lscape}

\def\about{$\sim$}
\def\chisq{$\chi^2$}
\def\chisqmin{$\chi^2_{min}$}
\def\e20{$\times 10^{20}$}
\def\emin12{$\times 10^{-12}$}

\def\>{$>$}
\def\<{$<$}
\def\sun{$_{\odot}$}

\def\ein{{\it Einstein}}
\def\newline{\hfil\break}
\def\mincir{\ \raise -2.truept\hbox{\rlap{\hbox{$\sim$}}\raise5.truept  
\hbox{$<$}\ }}                        
\def\magcir{\ \raise -2.truept\hbox{\rlap{\hbox{$\sim$}}\raise5.truept  
\hbox{$>$}\ }}                        
 
\def\bsax{BeppoSAX\,}
\def\degree{$^{\circ}$}
\def\ergcms{erg cm$^{-2}$ s$^{-1}$}
\def\ergs{erg s$^{-1}$}

\begin{document}

\thesaurus{11(11.05.1; 11.09.04; 11.09.1)} 

\title{ Broad band properties
  of medium and low L$\rm _x/L_b$ Early Type Galaxies}
  \author{ G. Trinchieri\inst{1}
\and 
S. Pellegrini\inst{2}
\and  
A. Wolter\inst{1}
\and 
G. Fabbiano\inst{3} 
\and 
F. Fiore\inst{4}
}
   \offprints{G.~Trinchieri}
   \mail{ginevra@brera.mi.astro.it}
  \institute{Osservatorio Astronomico di Brera, 
                 Via Brera 28, 20121 Milano ITALY
	\and Dipartimento di Astronomia, Universit\`a di Bologna,
             via Ranzani 1, 40127 Bologna, Italy
             \and Harvard-Smithsonian Center for Astrophysics 60 Garden St.,
                Cambridge, MA 02138 USA
	\and Osservatorio Astronomico di Roma, Via dell'Osservatorio,
	00044, Monteporzio Catone, Italy
}
 \date{ }
\titlerunning{Medium and low L$_x$/L$_b$ early-type galaxies}
 \maketitle
 \begin{abstract}
%
We have measured the spectral properties of five galaxies of low to
intermediate L$\rm _x/L_b$ ratios with \bsax\ and ASCA.  A hard component
(kT $\sim 4-10$ keV) is observed in all galaxies.  In NGC~1553 the
\bsax\ data show that this component is extended, and suggest an origin
for the emission in the evolved stellar population. 
In NGC~3115, a point-like source appears
embedded in an extended component morphologically similar to the stellar
body,  suggesting an almost equal contribution from the nuclear region
and the binary population.  A large central mass concentration and low
level optical activity in NGC~3379 argue for a contribution from the
nucleus also in this object.  However, for both NGC~3115 and NGC~3379,
the nuclear emission is at a level well below that
observed in other galaxies who host similar nuclear black holes. 
A second soft (kT $\sim$0.3- 0.7 keV) spectral component is needed to fit the
data of NGC~1407, NGC~1553 and NGC~4125 
over the entire energy range probed  (\about 0.2-10 keV), best
represented by a thermal component with line emission.  
We discuss possible interpretations of the origin of this component,
which however will be better defined only with
higher quality data. 
      \keywords{-- Galaxies: individual:
      NGC~1407, NGC~4125, NGC~1553, NGC~3379, NGC~3115
                -- Galaxies: elliptical and lenticular
                -- Galaxies: ISM }
   \end{abstract}

\section{Introduction}
\label{Intro}

One of the observational evidences from X--ray data of early type
galaxies is the large scatter in the correlation between the X--ray
and optical luminosities (Fabbiano, Kim and Trinchieri 1992): at a
given optical luminosity, the X--ray to optical flux ratios L$\rm _x/L_b$ may
range from those observed in the bulge of M31 and in spiral galaxies,
to values $\sim$50--100 times higher.  The presence of at least two
components, hot gas dominating at high X--ray luminosities, and the
evolved stellar population at faint X--ray luminosities, can explain
in part this observational evidence.  The spectral characteristics of the
emission also reflect the relative importance of these components:
already from the limited $Einstein$ spectral data, the 
average emission temperature is larger in galaxies with lower L$\rm _x/L_b$
(Kim et al. 1992), in agreement with the idea of an increasing
contribution of hard, individual X--ray sources relative to the hot
gas component as the L$\rm _x/L_b$ ratios decrease.
An additional very soft component was also found in the lowest L$\rm _x/L_b$
class, possibly due to a $\sim$0.2--0.4~keV interstellar medium or to
the collective emission of stellar sources (Pellegrini \& Fabbiano
1994).

\begin{table*}
\caption[] { General galaxy properties}
\begin{flushleft}
\begin{tabular}{ l l  l l  l  l l  l l  l }
\noalign{\smallskip}
\hline
\noalign{\smallskip}
Name & Type$^a$ & $B_{\rm T}^0\, ^a$ &Size $^a$  &  d$\, ^b$ & log$L_{\rm B}\, ^c$ & 
$\sigma \, ^d$ &  group$^e$        & group$^e$  \\
     &          &  (mag)             &(arcmins)  &  (Mpc)    & ($L_{\odot}$)       &
(km s$^{-1}$)  &  ($Einstein$) & ($ROSAT$)  \\
\noalign{\smallskip}
\hline
\noalign{\smallskip}
\hline
\noalign{\smallskip}
NGC1407 & E0 & 10.71 & $4.6\times 4.3$ &34.7 & 10.99 &  272 & 3 & 3  \\
NGC1553 & S0 & 10.26 & $4.5\times 2.8$ &21.5 & 10.75 &  200 & 3 & 2  \\
NGC3115 & S0 &  9.74 & $7.2\times 2.5$ &10.8 & 10.36 &  264 & 1 & 1  \\
NGC3379 & E1 & 10.18 & $5.4\times 4.8$ &13.0 & 10.35 &  209 & 1 & --  \\
NGC4125 & E6 & 10.67 & $5.8\times 3.2$ &38.0 & 11.08 &  232 & 1-2 & 2  \\
\noalign{\smallskip}
\hline
\end{tabular} 
\end{flushleft}
\bigskip

$^a$ from de Vaucouleurs et al. 1991. 
$B_{\rm T}^0$ is the total B magnitude, corrected for 
Galactic and internal extinction; the size gives the apparent major
and minor axes diameters at the surface brightness level of 25 mag/square
arcseconds.

$^b$ distance from Fabbiano et al. (1992), who adopt a Hubble constant of 50 
km s$^{-1}$ Mpc$^{-1}$.

$^c$ total B-band luminosity $L_B$, derived using the indicated distance and 
$B_{\rm T}^0$. 

$^d$ central stellar velocity dispersion from McElroy (1995). 

$^e$ Group into which the galaxy has been classified on the basis of
its L(0.2--4 keV)/L$_{\rm B}$ ratio (for $Einstein$ data, Kim et
al. 1992), or its L(0.5--2.0 keV)/L$_{\rm B}$ ratio (for $ROSAT$ data,
Irwin \& Sarazin 1998). The X-ray faintest galaxies belong to group 1
while the X-ray brightest to group 4. The group boundaries differ in
the two papers; in both works they have been chosen so that each
group contains roughly the same number of galaxies of the same general
spectral characteristics.

\label{gen}
\end{table*}

The ROSAT PSPC and ASCA data confirm
the \ein\ results: a soft (kT$\sim$0.5-1~keV) optically thin emission due
to hot gas dominates 
in X-ray bright objects, 
and an harder component, with kT $\ge 3-4$ keV, 
most likely associated with the evolved stellar population 
(Matsushita et al. 1994; Matsumoto et
al. 1997; Buote 1999), or to nuclear activity 
(Allen, Di Matteo \& Fabian 2000; Matsumoto et al.  1997) and present in
all galaxies, dominates 
in X-ray faint objects.
In the lowest L$\rm _x/L_b$ class, 
a very soft component is also measured, with a temperature now  well
constrained to $\sim$0.2-0.3 keV 
(Fabbiano et al 1994; Pellegrini 1994; Fabbiano \& Schweizer
1995; Kim et al 1996).  However, the origin of the very soft component
has not been properly understood yet: it could be
placed in stellar sources, in X-ray binaries (Irwin
\& Sarazin 1998), or could be a cooler phase of the ISM: 
as detailed hydrodynamical
simulations show (Pellegrini \& Fabbiano 1994), hot gas with the
required emission temperature and luminosity can be retained by the
galaxies.  However, the hot gas temperature is comparable to that of the very
soft component only in galaxies with quite shallow potential wells,
i.e., with central stellar velocity dispersion $\sigma \leq 200$ km s$^{-1}$.

More galaxies with intermediate to low L$\rm _x/L_b$ ratios still need to be
investigated. We report here the results on the \bsax\ and ASCA 
observations of 5 such galaxies.  Their 
general optical properties are summarized in Table~\ref{gen} together
with the $\rm L_X / L_B$ group in which they fall according to their 
soft band X--ray fluxes. 
The wide energy band and the good
spectral resolution of both satellites are suitable to detect and
measure separately the amount and spectral parameters of the different
emission components.  Moreover, the spatial resolution of \bsax\ at
high energies
allows us to study the  spatial characteristics of the hard component,
to determine its extent. 
We discuss the observations and data analysis in \S 2-6, we briefly
compare the spectral results with those reported in the literature from
previous
missions and/or different authors in \S 7,  and we discuss the
results in \S 9.   In \S 8 we further discuss a detection at very high energy
of a source in the field of NGC~1553, but most likely 
unrelated to this galaxy.

\begin{table}
\caption[]{Log of the Observations}
\bigskip

\begin{flushleft}
\begin{tabular}{llrr} \hline
Object&Instrum  & Obs.  Dates & On Time    \\
&& Beginning - End &\multicolumn{1}{c}{(s)} \\
\hline
NGC 1407&GIS2&23/08/95 - 24/08/95&34759  \\
&GIS3&23/08/95 - 24/08/95 &34755  \\
&SIS0&23/08/95 - 24/08/95 &33801\\
&SIS1&23/08/95 - 23/08/95 &33747\\
\hline
\hline
NGC 1553& MECS1& 16/01/97 - 17/01/97& 25880 \\
&MECS2&16/01/97 - 17/01/97& 25849 \\
&&16/11/97 - 17/11/97& 30019 \\
&MECS3& 16/01/97 - 17/01/97& 25664 \\
&&16/11/97 - 17/11/97&  29962 \\
&LECS & 16/01/97 - 17/01/97& 14262 \\
&     & 16/11/97 - 17/11/97&18763 \\
\hline
\hline
NGC 3115&MECS2 & 15/05/98 - 16/05/98 & 35986 \\ 
&MECS3 & 15/05/98 -  16/05/98 & 35952 \\
  &MECS2 &    26/05/98 -   27/05/98 &36594 \\
  &MECS3 &    26/05/98 -   27/05/98 &36543\\
&LECS &15/05/98 - 16/05/98 &14997 \\
&LECS &26/05/98 - 27/05/98&13177 \\
\hline
\hline
NGC 3379 &MECS2 & 14/12/98 - 17/12/98 & 99122 \\
&MECS3 &    14/12/98 - 17/12/98 &98749 \\
&LECS & 14/12/98 - 17/12/98 &34594 \\
\hline
\hline
NGC 4125& MECS1& 26/04/97 - 27/04/97& 57128 \\
&MECS2 &  26/04/97 - 27/04/97& 57148\\
& MECS3 &  26/04/97 - 27/04/97& 56971 \\
& LECS &  26/04/97 - 27/04/97& 22189 \\
\hline
&GIS2&05/04/95 - 05/04/95 &38634  \\
&GIS3&05/04/95 - 05/04/95&38634  \\
&SIS0&05/04/95 - 05/04/95&30073\\
&SIS1&05/04/95 - 05/04/95 &21505 \\
\end{tabular}
\end{flushleft}
\label{log}
\end{table}

\section{Data Analysis }
\label{data}

Five galaxies have been observed with ASCA and \bsax, as summarized in
Table~\ref{log}.  While the details of the data analysis are different
for each instrument and to a certain degree for each observation, we
have followed a similar plan for all.  We have first produced smoothed
images in the broadest energy band compatible with the statistical
significance of the image and with the characteristics of the
different instruments (shown in figures~\ref{n1553-map}, ~\ref{n3115-map},
~\ref{n3379-map}, ~\ref{n4125-map} and ~\ref{n1407-map}
superposed onto the optical images from the Digitized Sky Survey
plates\footnote{The Digitized Sky Survey was produced at the Space
Telescope Science Institute (STScI) under U.S. Government grant NAG
W-2166.}).  We have then determined the extent of the source by
comparing the radial profiles of the total emission with the expected
background profile, taking into account possible asymmetries in the source
morphology, either evident from the present data or with the help of
higher resolution ROSAT images.  We have then produced the spectral
files, both for the source and its background.
We have used the latest release of standard software packages 
to produce images, spectral photon distributions
and ancillary spectral files from the event files, and
\verb+IRAF+/PROS  and XSPEC for the spatial and spectral analysis, 
respectively.
The details of the procedures used for each galaxy are
given below.

\section{BeppoSAX data}
\label{datasax}

A comprehensive description of the satellite and of the instruments
operating on it can be found in Boella et al. (1997a and b);  Parmar et al.
1997; Frontera et al. (1997), Manzo et al. (1997) and references
therein. 

We have used the two imaging instruments 
on board of \bsax: 
the Low Energy concentrator spectrometer  (LECS), operating
in the $\sim$ 0.1-10 keV band, with a circular Field of View (FoV) of $
\sim 18\farcm5$ radius;  and the Medium Energy concentrator spectrometer
(MECS), consisting of 3
identical active units\footnote{In May 1997 detector unit 1 of the
MECS developed a fault.  Since then only units 2 and 3 are 
operating.}, 
operating in the $\sim 1.5-10$  keV band with a $\sim 28'$ radius FoV.
We also report on a detection of emission at higher energies
during the observation of NGC~1553 with the  Phoswich Detector
System (PDS), sensitive in the $\sim 15-300$ keV band and 
co-aligned with LECS and MECS.

As shown in Table~\ref{log}, 
NGC~1553 was observed in two separate occasions, about one year apart. The
length of each observation is similar, but only two MECS were operating
at the time of the second observation, which reflects in the smaller
statistics obtained.  Each observation is therefore treated
independently.  The observation of NGC~3115 is also split into two
segments;  however, they are close in time and the same set of
instruments is operating, therefore we have merged the two segments in
a single set of data.  Only the observation of NGC~4125 was completed
with all 3 MECS instruments operating.

We have used the event files provided by the SAX Data Center (SDC) that
have already been screened with standard criteria, and the 
blank sky event files as background templates, available from the
anonymous ftp SAX-SDC site, that have been also been processed with the
same criteria.

\begin{figure*}
\unitlength1.0cm
\begin{picture}(18,9.5)
\thicklines
\put(0,0.0){
\begin{picture}(18,9.0)
\resizebox{18cm}{!}{
\psfig{figure=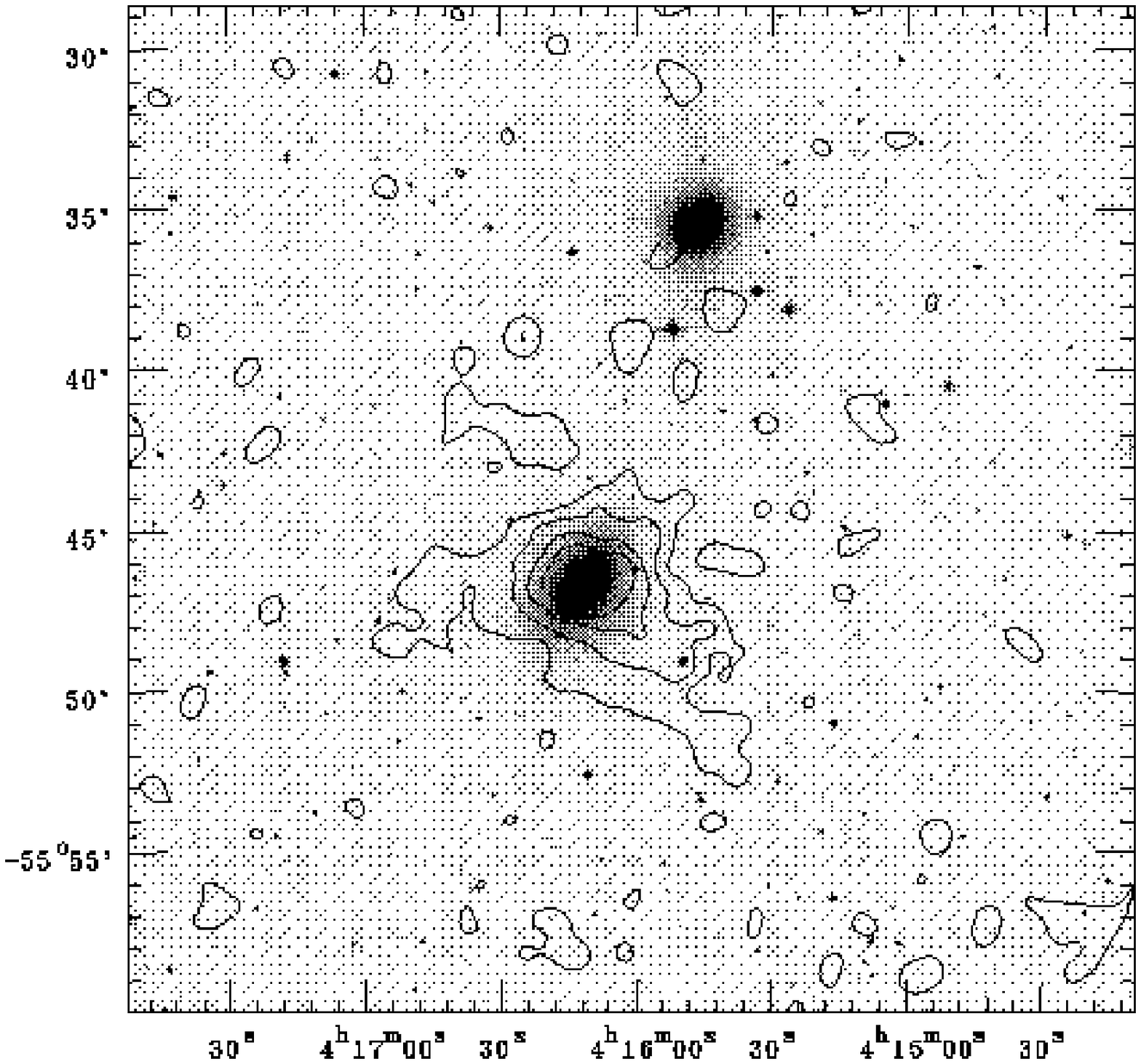,width=18cm,clip=}
\psfig{figure=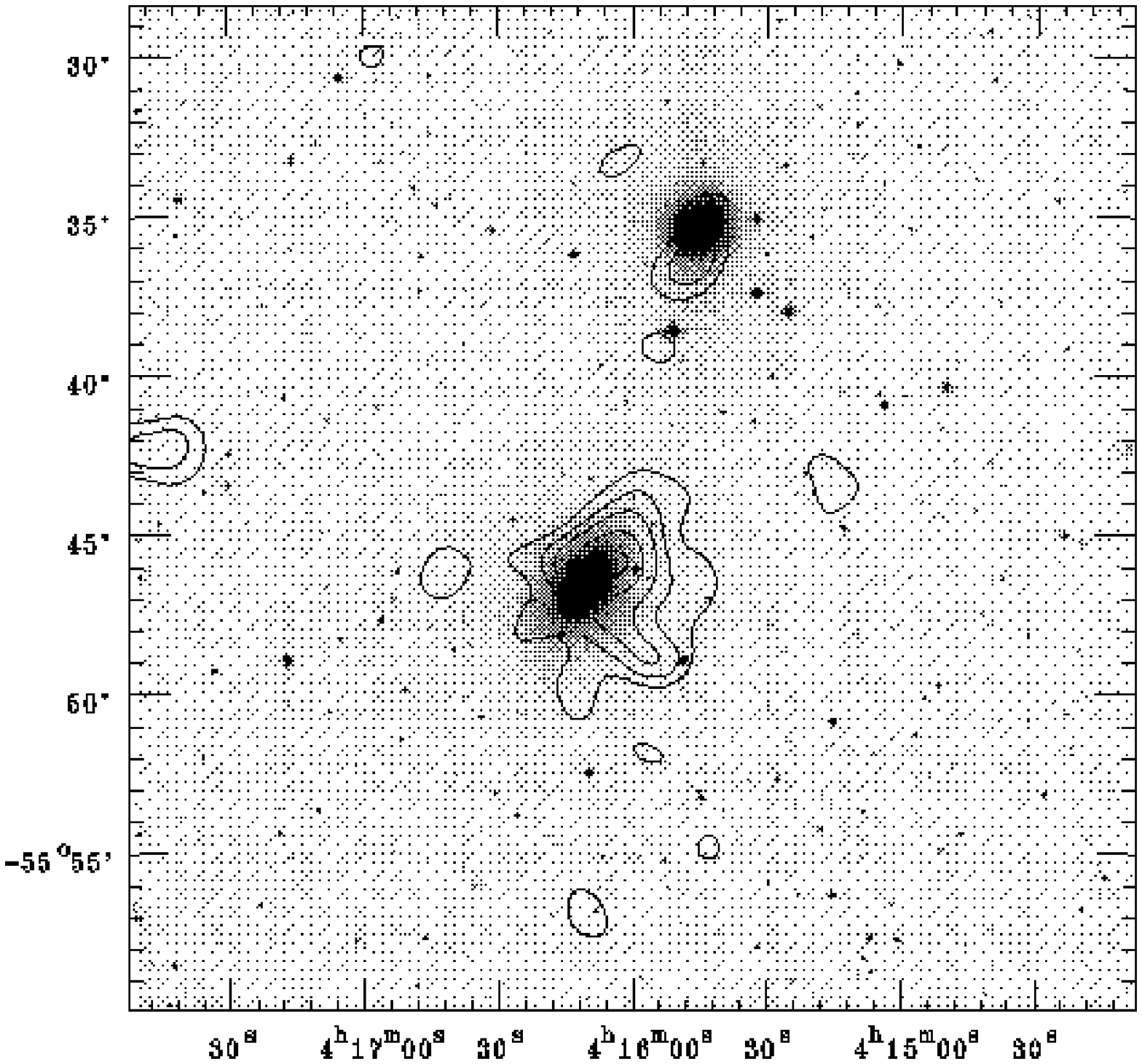,width=18cm,clip=}
}
\end{picture}}
\put(3.4,8.7){
NGC~1553-MECS
}
\put(4.5,6.8){
NGC~1549
}
\put(2.3,4){
NGC~1553}
\put(12.4,8.7){
NGC~1553-LECS}
\put(13.5,6.8){
NGC~1549
}
\put(11.3,4){
NGC~1553}
\end{picture}
\caption[]{Contour plots of the SAX images of NGC~1553.  LEFT: MECS data, in
the 2-10 keV range, smoothed  with a Gaussian
function with $\sigma$=$24''$.  RIGHT: LECS data, in the 0.1-5 keV range, 
smoothed with a Gaussian function with $\sigma$=$32''$.
\label{n1553-map}
 }
\end{figure*}

\begin{figure*}
\unitlength1.0cm
\begin{picture}(18,9.5)
\thicklines
\put(0,0.0){
\begin{picture}(18,9.0)
\resizebox{18cm}{!}{
\psfig{figure=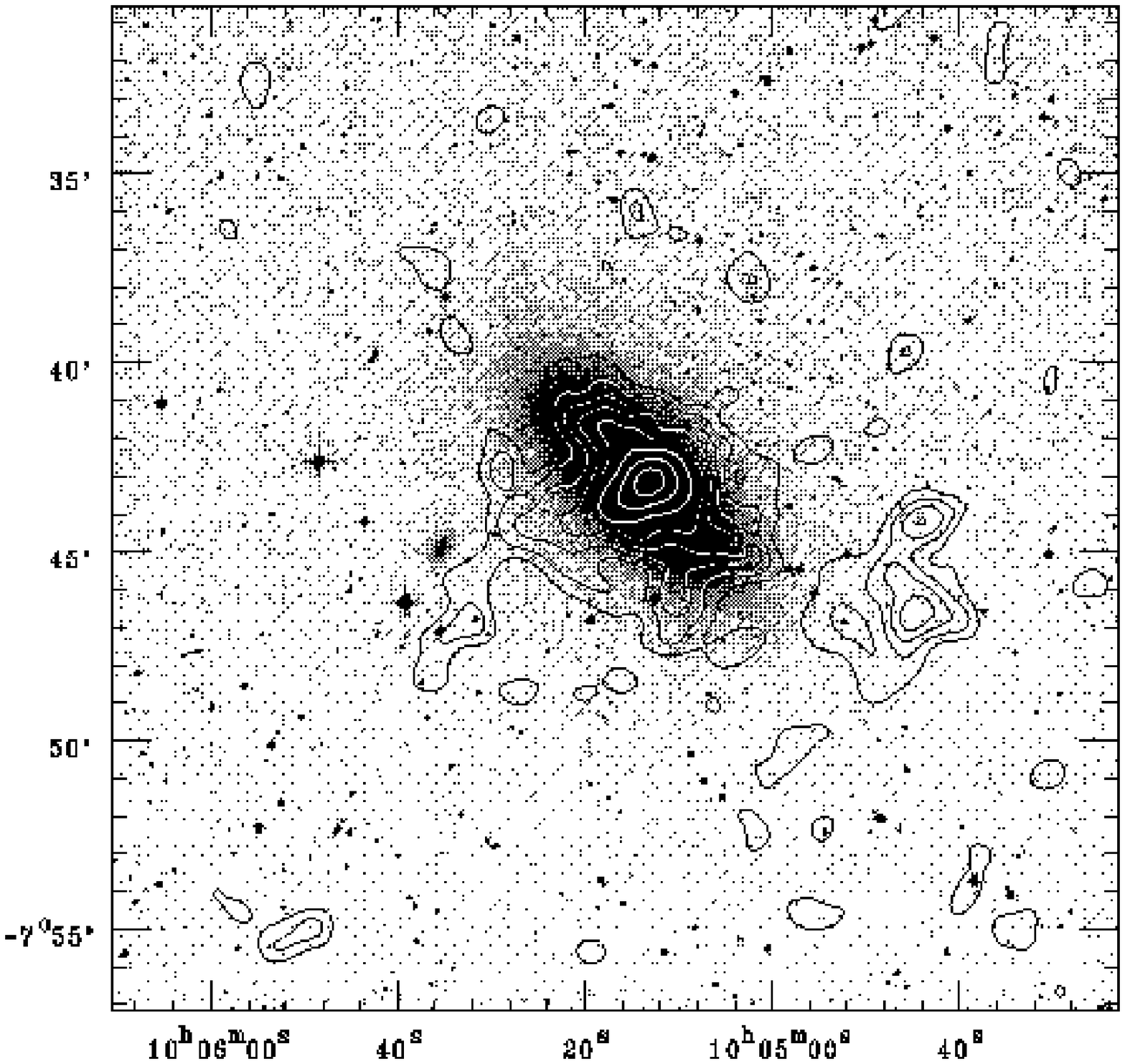,width=18cm,clip=}
\psfig{figure=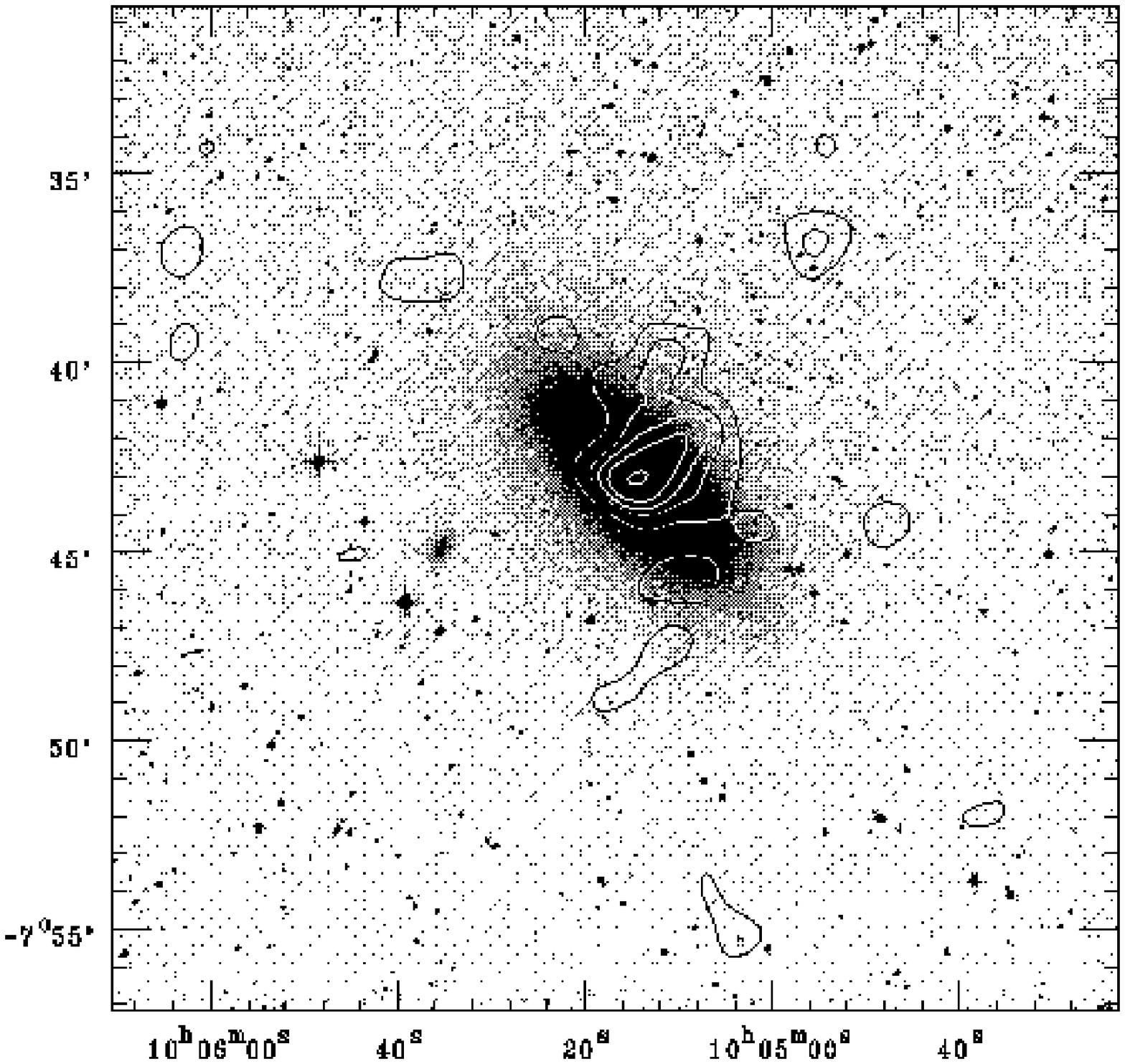,width=18cm,clip=}
}
\end{picture}}
\put(3.4,8.7){
NGC~3115-MECS
}
\put(12.4,8.7){
NGC~3115-LECS}
\end{picture}
\caption[]{As in Fig.~\ref{n1553-map} for NGC~3115.
 }
\label{n3115-map}
\end{figure*}

\begin{figure*}
\unitlength1.0cm
\begin{picture}(18,9.5)
\thicklines
\put(0,0.0){
\begin{picture}(18,9.0)
\resizebox{18cm}{!}{
\psfig{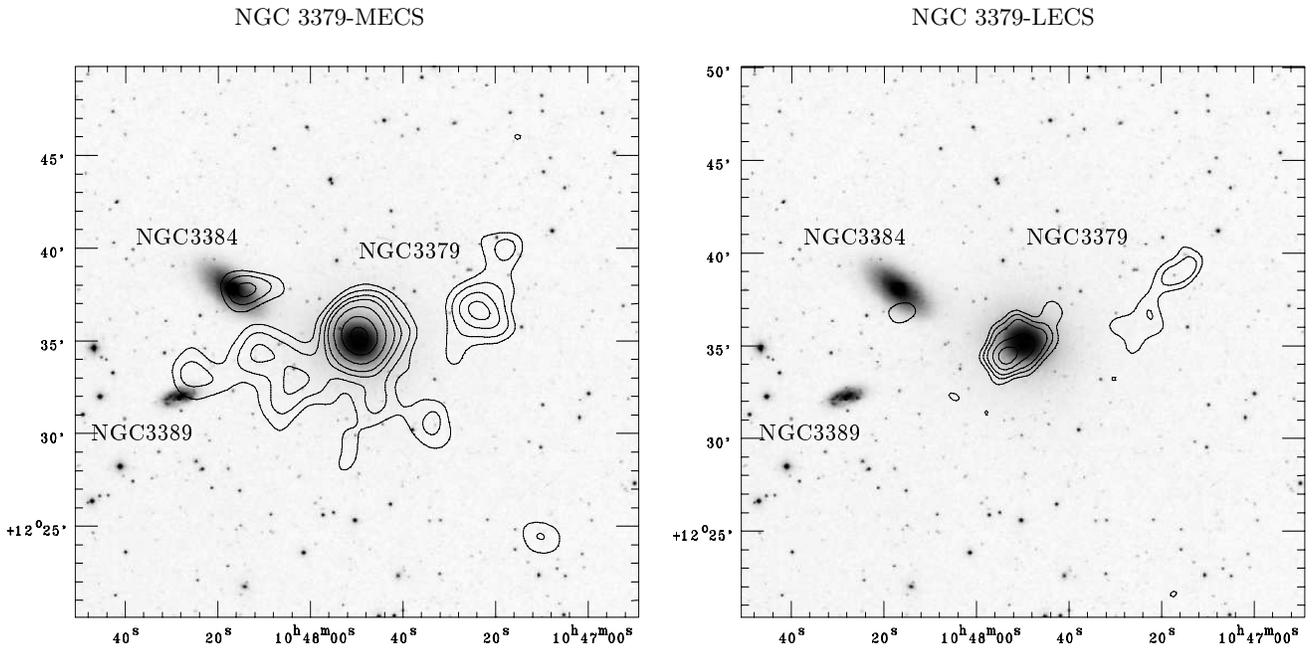}
}
\end{picture}}
\put(3.4,8.7){
NGC~3379-MECS
}
\put(12.4,8.7){
NGC~3379-LECS}
\end{picture}
\caption[]{As in Fig.~\ref{n1553-map} for NGC~3379.
 }
\label{n3379-map}
\end{figure*}

\begin{figure*}
\unitlength1.0cm
\begin{picture}(18,9.5)
\thicklines
\put(0,0.0){
\begin{picture}(18,9.0)
\resizebox{18cm}{!}{
\psfig{figure=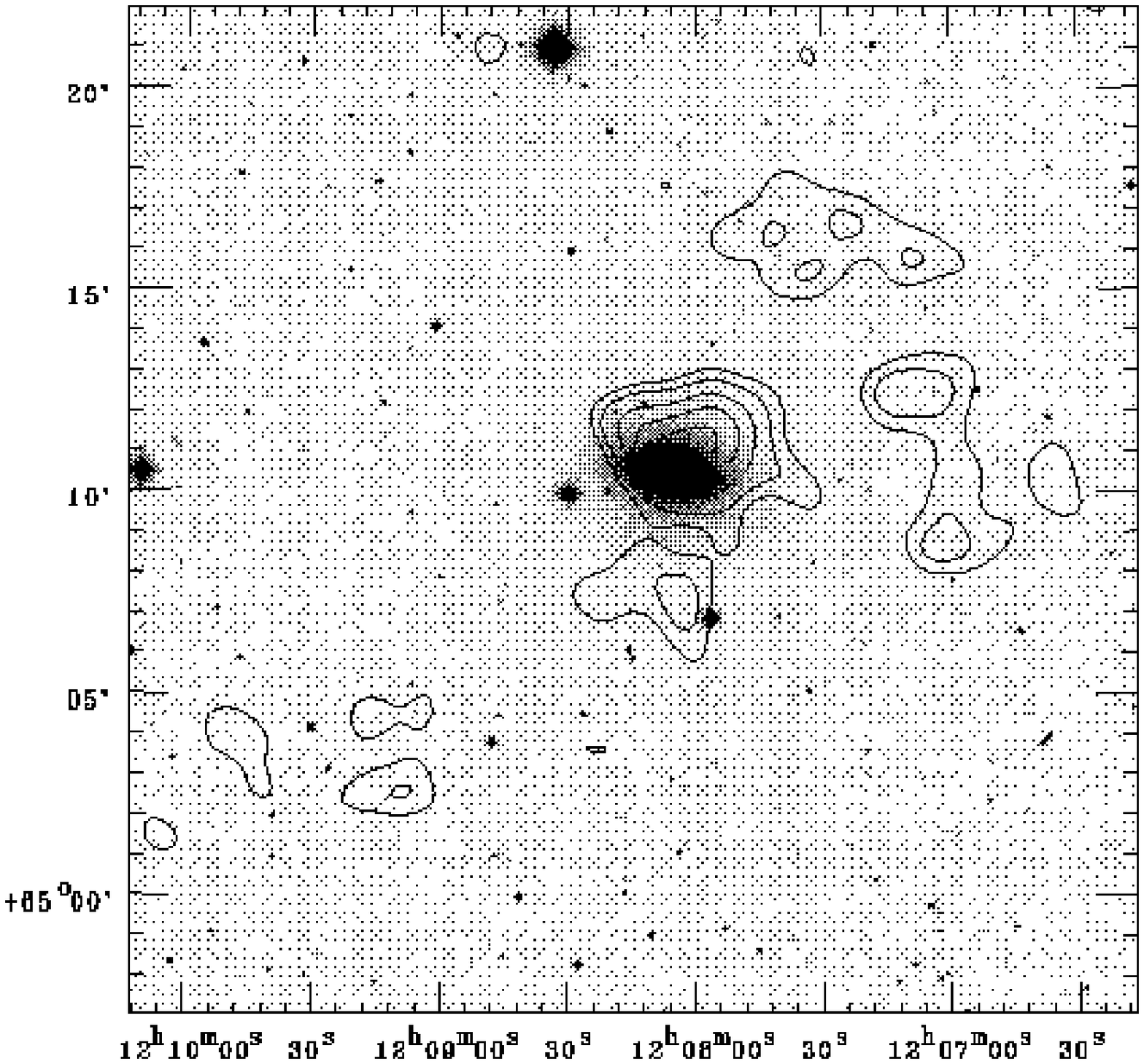,width=18cm,clip=}
\psfig{figure=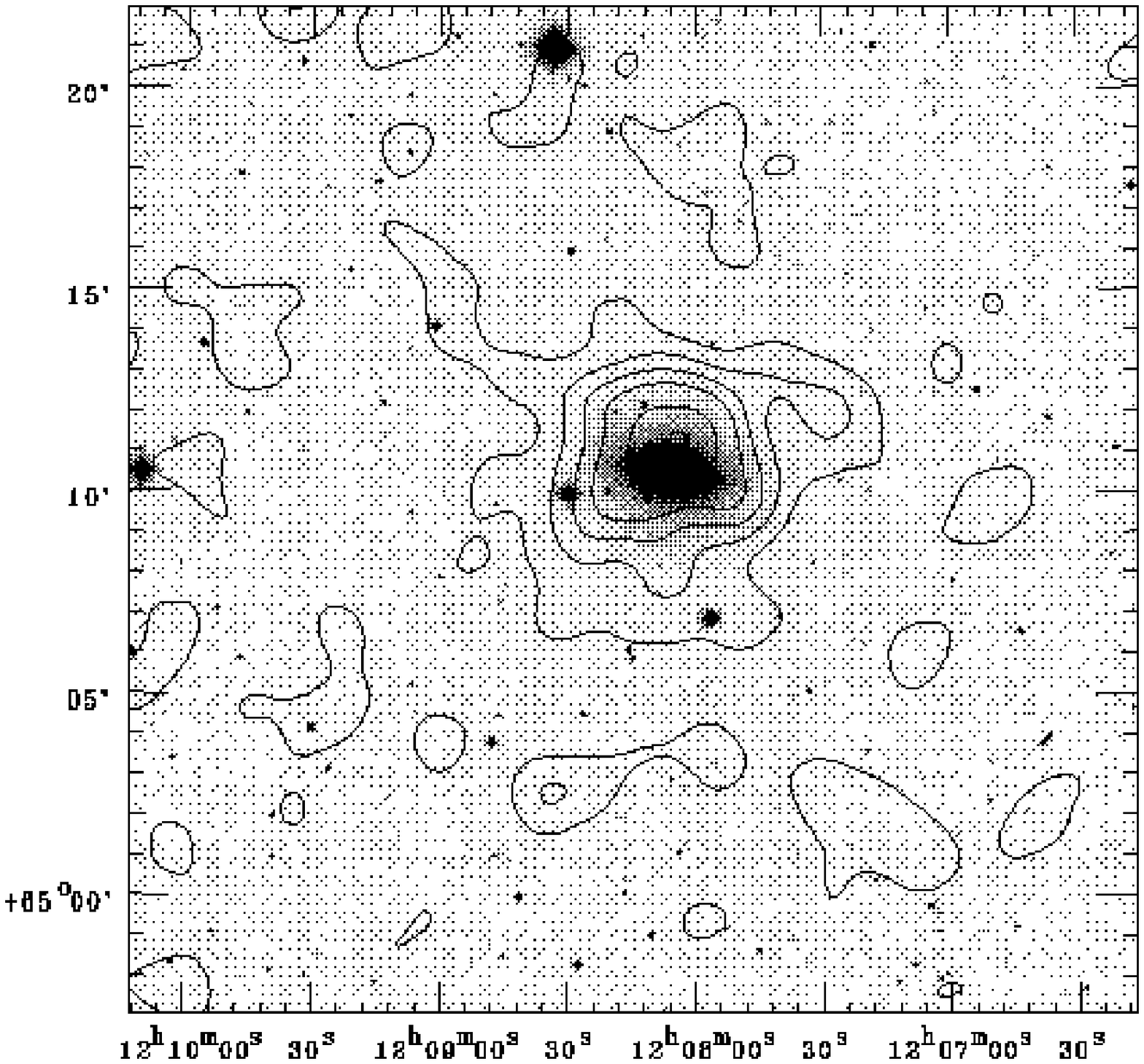,width=18cm,clip=}
}
\end{picture}}
\put(3.4,8.7){
NGC~4125-MECS
}
\put(12.4,8.7){
NGC~4125-LECS}
\end{picture}
\caption[]{As in Fig.~\ref{n1553-map} for NGC~4125.
 }
\label{n4125-map}
\end{figure*}

Figures~\ref{n1553-map}, \ref{n3115-map}, \ref{n3379-map} and
\ref{n4125-map} show the smoothed images obtained from  MECS and LECS
observations for each object superposed onto the optical images.   A
comparison between the BeppoSAX positions (of the target galaxy and of
other sources in the field) and  both the optical images and the ROSAT
images shows an agreement to within \< 30$''$, consistent with the
expected BeppoSAX positional error (see Fiore et al. 2000).   The possible 
off-set between LECS and MECS images is also within expectations (see 
Trinchieri et al. 1999).

The radial profiles of the total emission observed in the MECS
instruments, azimuthally averaged in concentric annuli about the X--ray
peak, are shown in Fig.~\ref{raw}.  In all cases a radial decrease of
the emission is observed, followed by a flattening of the profiles
at radii in the range $\sim 4'-8'$, outside of which the
profiles are similar to the profiles from the blank sky
fields.  These have been obtained in the same regions and rescaled for
the relative exposure times.   In a couple of cases (shown in
Fig.~\ref{raw}), the rescaling factors had to be corrected (by $\sim 10$\%)
to reproduce the level of the background emission in the field.

Figure~\ref{net} shows the comparison of the net emission from the
target galaxies detected in the $\sim 2-10$ keV band  with the MECS
Point Spread Function (PSF).  The PSF can be described analytically as
a function of energy (Boella et al.  1997b), however since it depends on
a combination of the effects of the mirror and of the instrument
responses, it appears almost independent of energy above a few keV.  We
have therefore chosen to compare the MECS profiles both with the
``worst case" PSF, namely at the softest energy in our band (2 keV),
and at an intermediate energy (5 keV).

The LECS data, that extend the energy coverage to lower energies,
have a smaller statistical significance due to both the much
shorter exposure times (the instrument is operated only
during satellite dark time), and the fact that there is only one instrument
operating (instead of 2-3).   Consequently a  detailed spatial
analysis is in most cases not feasible.  We have therefore disregarded
the spatial information from the  LECS data, and  produced
radial profiles only to check the level  of the background
against the blank sky profiles.  For the
spectral analysis, we have typically used the same size region
determined in the MECS data, centered on the source peak.

\subsection{NGC~1553}

The source position is $\sim 2\farcm6$ off-axis, since we had tried to
include NGC~1549 as well in the field of view.  Unfortunately, NGC~1549
is too weak to be detected within this observation with the MECS.  A
possible enhancement is observed with the LECS at the position of this
source, mostly at soft energies, but not with sufficient statistical
significance.   The off-axis position of NGC~1553 is such that
there is no significant degrading of the instrument performances.  It
has however a significant impact in the handling of the data analysis,
mostly in the determination of the background, which is not uniform
across the detector\footnote{ This is due to the fact that the fluorescence
emitted from the two calibration sources located at opposite sides of
the detector is probably reabsorbed, and results in enhancement in the
internal background near the position of the sources and in the region
of the detector that connects the two sources.}.  Therefore, the
background for off-axis sources needs to be estimated at the same
position in the detector to better account for the additional
background produced by these calibration sources (for details, see the
``Cookbook" available on line at the SAX-SDC site).  Moreover, the 
source is at different detector positions in the three MECS units.
To produce the
radial profile of the background shown in Fig.~\ref{raw}, we have
summed the azimuthally averaged surface brightness obtained
separately for each MECS in concentric annuli, centered as close as
possible to the position of the X--ray peak in detector coordinates.

\begin{figure*}
\unitlength1.0cm
\begin{picture}(18,9.5)
\thicklines
\put(0,0.0){
\begin{picture}(18,9.0)
\resizebox{18cm}{!}{
\psfig{figure=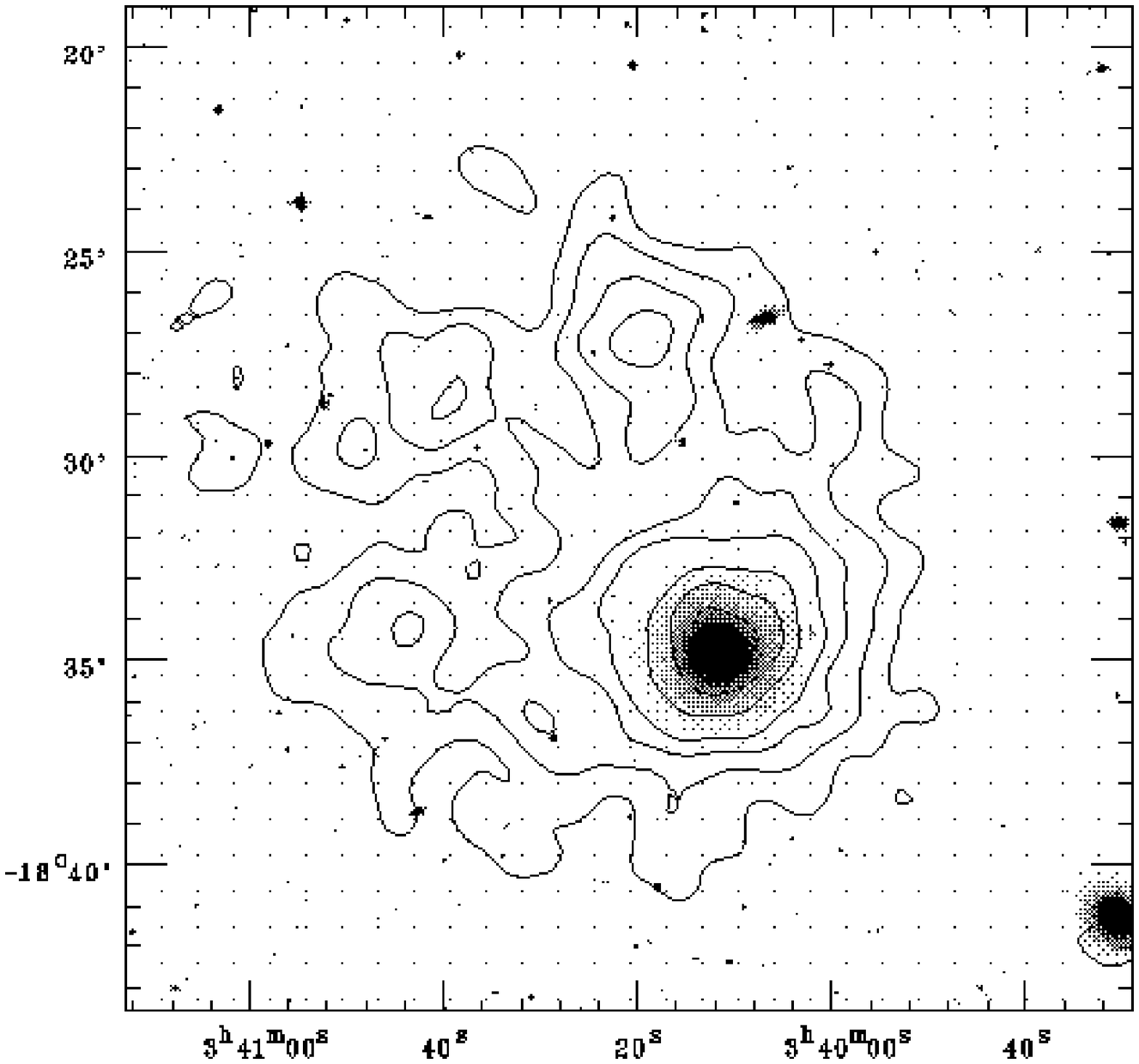,width=16cm,clip=} 
\psfig{figure=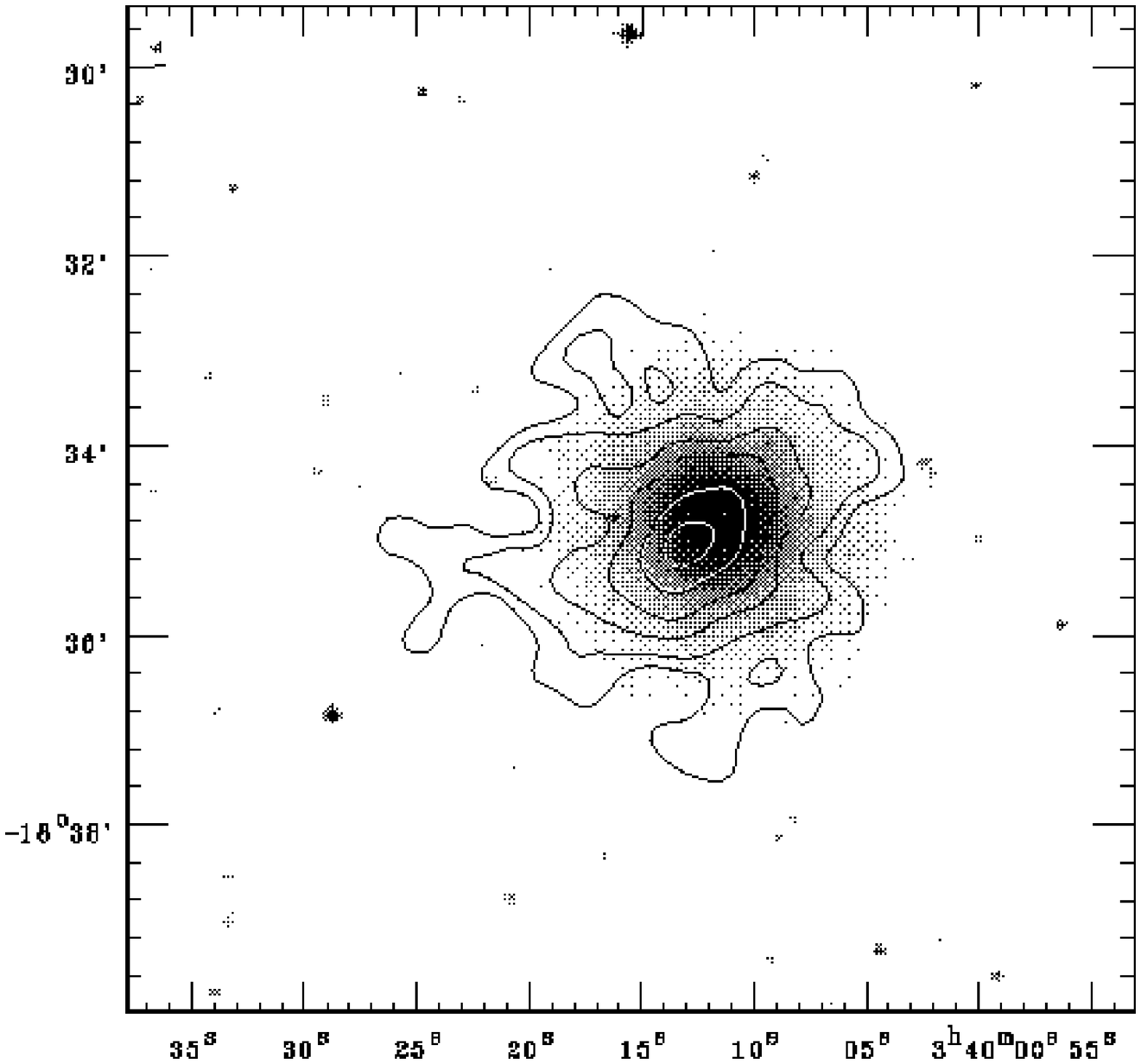,width=16cm,clip=} 
} 
\end{picture}}
\put(3.4,8.7){
NGC~1407-GIS
}
\put(12.4,8.7){
NGC~1407-SIS}
\end{picture}
\caption[]{Contour plots of the GIS (left) and SIS (right) 
images of NGC~1407.  The data have been smoothed with a Gaussian
function with $\sigma$=$13''$. }
\label{n1407-map}
\end{figure*}

\begin{figure*}
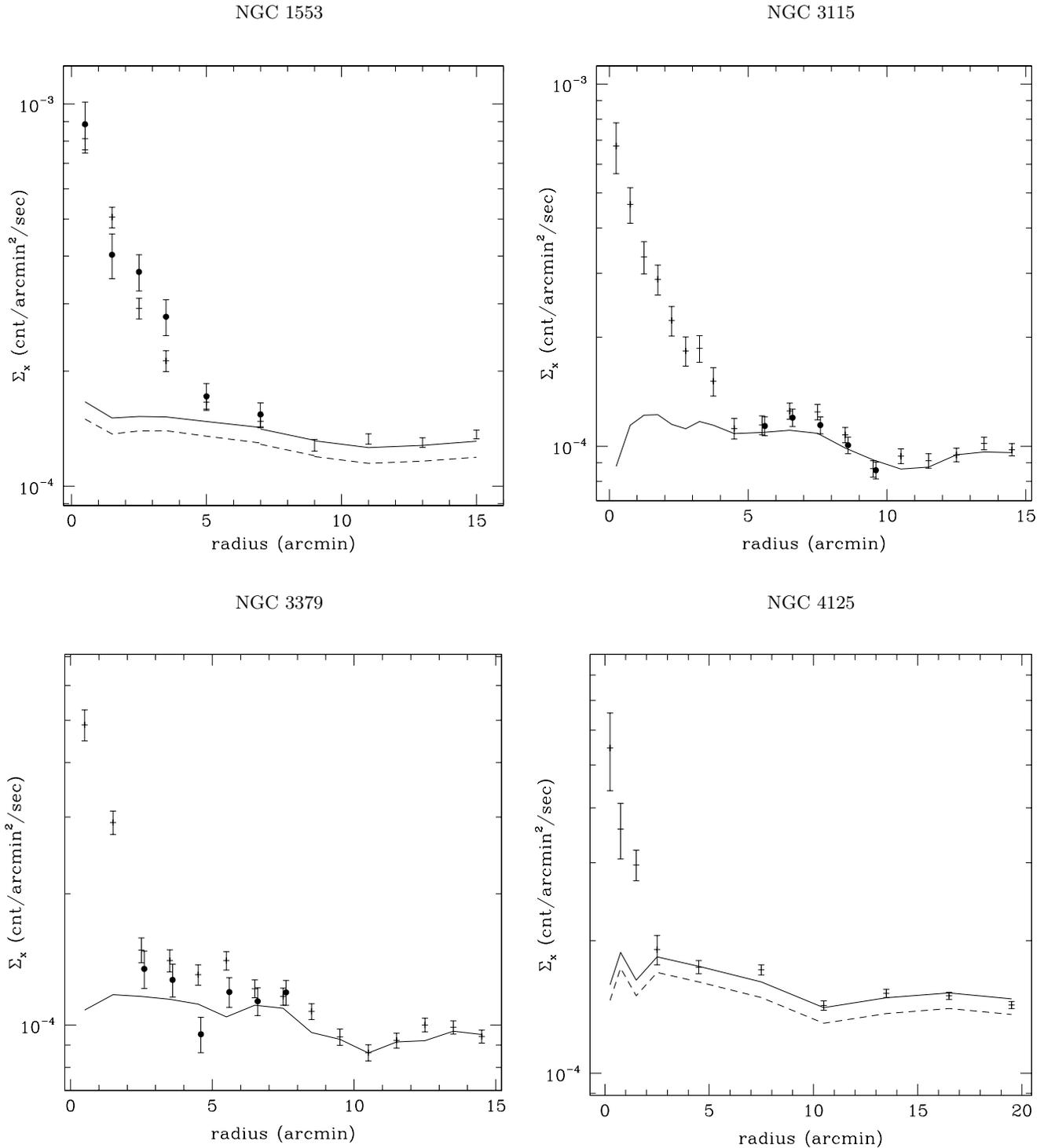

\unitlength1.0cm
\begin{picture}(18,20.0)
\thicklines
\put(0,10.0){
\begin{picture}(18,9.0)
\resizebox{18cm}{!}{
{
\psfig{figure=ms10097.f6a,width=16cm,clip=}
\psfig{figure=ms10097.f6b,width=16cm,clip=}
}}
\end{picture}}
\put(0,0){
\begin{picture}(18,9.0)
\resizebox{18cm}{!}{
\psfig{figure=ms10097.f6c,width=16cm,clip=}
\psfig{figure=ms10097.f6d,width=16cm,clip=}
}
\end{picture}}
\put(4.0,9.2){
NGC~3379}
\put(13,9.2){
NGC~4125
}
\put(4.0,19.2){
NGC~1553}
\put(13,19.2){
NGC~3115
}
\end{picture}

\caption[]
{\footnotesize{Radial profile of the total emission (symbols plus error
bars) compared to the emission from the blank sky fields (solid line) from the
MECS data.
The profiles are obtained in concentric adjacent annuli about the
X--ray peak and azimuthally averaged, unless otherwise noticed.  For
NGC~1553 and NGC~4125, two background levels, from the blank sky fields
differently normalized to the data, are shown (see text for details).  The solid
line represents the level assumed.
{\bf NGC~1553}:  The profiles are obtained in each detector separately
and added together. The background profile is centered at the same
detector position of the source.  Dots: azimuthal angles
270\degree-180\degree.
Angles are counterclockwise from North.  
{\bf NGC~3115}: Filled dots
indicate the profile obtained by masking out a circle of $\sim 2\farcm3$
radius at the position of the SW source (see text).  The points are
slightly offset in radius for clarity.  
{\bf NGC~3379}: Filled dots
indicate the profile obtained in the Northern half only.  A slight
offset in radius is added for clarity.  In both profiles, the source to
the NW is masked out with a circle of radius r$\sim 2'$.  {\bf
NGC~4125}:  two levels for the background profile are shown (see text). 
}}
\label{raw}
\end{figure*}

\begin{figure*}
\psfig{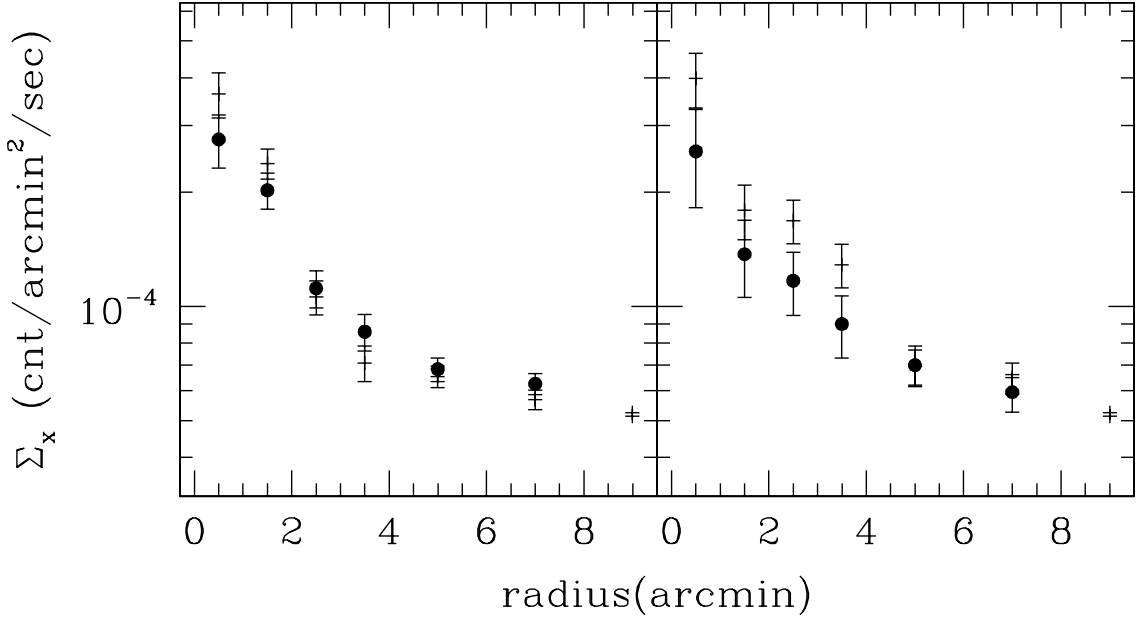}
\caption[]{Comparison of the radial profiles of the total emission in the MECS 
from different azimuthal sectors in the two observations of NGC~1553.  
LEFT: angles 
-90\degree to 180\degree are averaged together.  Right: angles 180\degree 
to 270\degree are averaged
together.  $+$ signs are for the first observation, filled circles are for the
second observation.}
\label{n1553-var}
\end{figure*}

ROSAT data of this galaxy show that the source is extended, and that
the emission in a radius of $\sim 5'$ has a general elongation in the
NW-SE direction (Trinchieri et al. 1997).  A separate source is visible
at $\sim 3\farcm6$ to the SW of the galaxy.  The BeppoSAX image also
suggests an extended source with a clear elongation in the SW direction,
$\sim 4'$ long, visible both in the LECS and in the MECS images.    The
azimuthally averaged surface brightness ($+$ signs) decreases out to a
radius of $\sim 8'$ and then flattens.  However, a significant fraction
of the photons  outside of r$\sim 2'$ are located in the SW quadrant,
presumably  due to the source detected with the ROSAT HRI.  This is
further confirmed by the fact that the comparison between the profiles
in the SW quadrants in the first and the second observations  suggest
that the source has varied in intensity (fainter at the time of the second
observation, see Fig.~\ref{n1553-var}).

\begin{figure*}
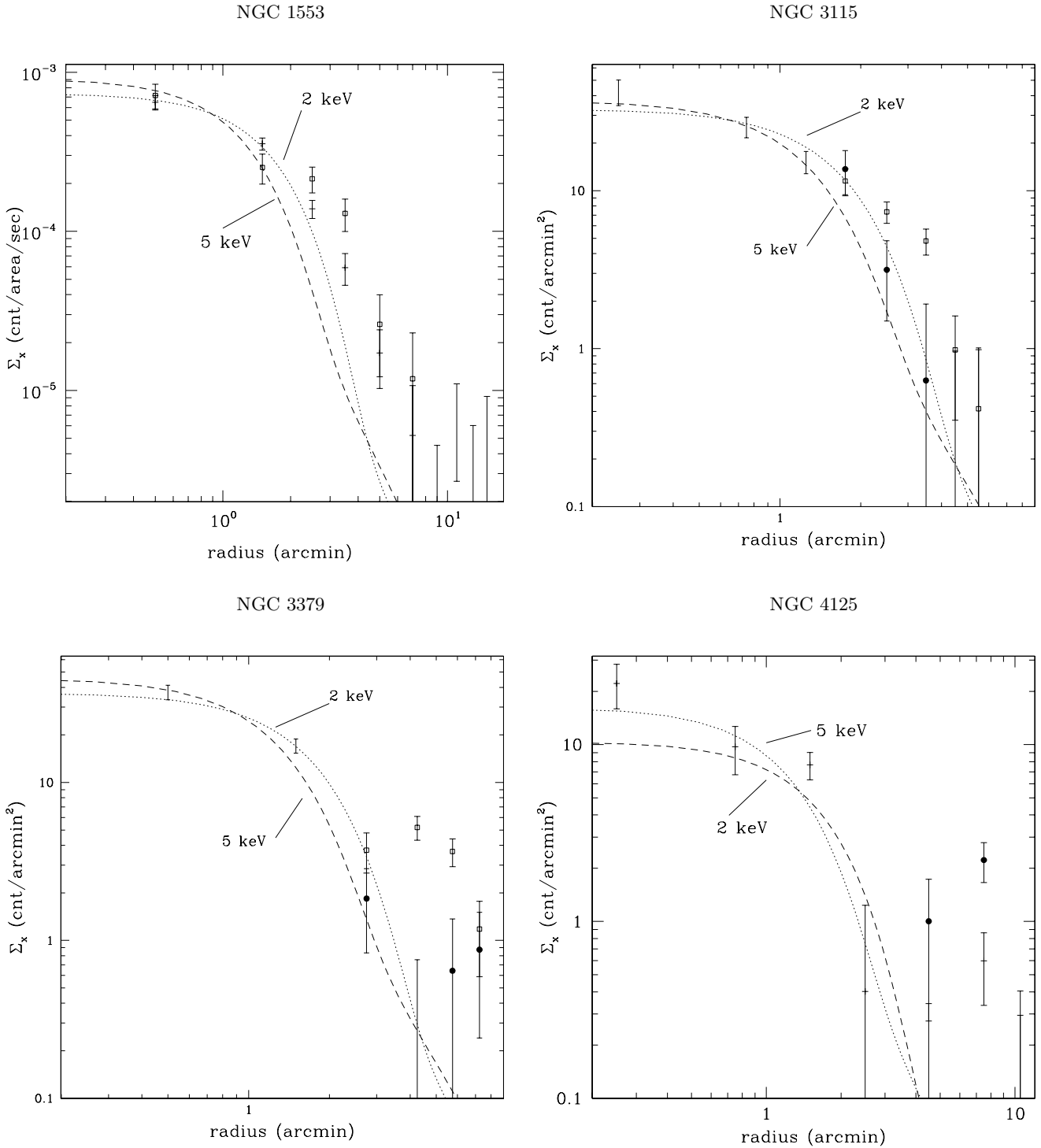

\unitlength1.0cm
\begin{picture}(18,20.0)
\thicklines
\put(0,10.0){
\begin{picture}(18,9.0)
\resizebox{18cm}{!}{
\psfig{figure=ms10097.f8a,width=16cm,clip=}
\psfig{figure=ms10097.f8b,width=16cm,clip=}
}
\end{picture}}
\put(0,0){
\begin{picture}(18,9.0)
\resizebox{18cm}{!}{
\psfig{figure=ms10097.f8c,width=16cm,clip=}
\psfig{figure=ms10097.f8d,width=16cm,clip=}
}
\end{picture}}
\put(4.0,9.2){
NGC~3379}
\put(13,9.2){
NGC~4125
}
\put(4.0,19.2){
NGC~1553}
\put(13,19.2){
NGC~3115
}
\end{picture}
\caption[]{\footnotesize{Radial distribution of the net profiles for the four galaxies
(symbols with error bars) detected with the MECS.  
The profile from the  PSF at two energies (dashed and dotted lines) is 
arbitrarily normalized to the data.
{\bf NGC~1553}:  the profile is obtained in different angular sectors.    
$+$ symbols: azimuthal angles 180\degr-270\degr; squares: 
azimuthal angles 270\degr-180\degr.  Angles
are counterclockwise from North.  Azimuthal averages outside of 9$'$. 
{\bf NGC~3115}: The source to the SW (see text) is masked out with 
a circle of $\sim 2\farcm3$ radius.  Data within $1\farcm5$ are azimuthally averaged.
Symbols outside $1\farcm5$ indicate: 
azimuthal angles 0\degr-270\degr (boxes); azimuthal angles 
270\degr-360\degr (filled dots).
{\bf NGC~3379}: The profile within a radius of $2'$ is azimuthally averaged. 
At larger radii, the profile is obtained in the Southern
(open boxes) and Northern (filled dots) halves. 
The
source to the NW is masked out with a circle of radius r$\sim 2'$ (see text). 
{\bf NGC~4125}:
$+$ symbols indicate the azimuthally averaged profile.  Filled dots
refer to the NW sector only.}
}
\label{net}
\end{figure*}

The comparison between the distribution of the source photons and the
\bsax PSF (Fig~\ref{net}) however confirms the extent of the
source, out to a radius r$\sim 6'-8'$: the source profile is
inconsistent with the PSF at all energies also for azimuthal angles
270\degree $-$180\degree ($i.e.$ excluding the SW quadrant). 

For the spectral analysis we have therefore considered  a source region
of $6'$ radius, excluding a circular region of $3'$ at the peak of the
SW source (estimated from the ROSAT image as an off-set from the
galaxy's peak).  While significantly reducing the  contamination, we
are left with a peculiarly shaped region, that cannot be properly
accounted for to compute the fluxes and similar 
quantities.  However, this should not introduce a bias in the
spectral analysis:  the files used in
the  spectral fits that account for photons distributed outside of the
source region assume a point source distribution of the photons, which
provides an approximation for the true distribution, and we have used a
relatively large radius, so that most of the photons are included in
the extraction region.   The flux however will need to be properly corrected
(see \S~\ref{fluxes}).

\subsection{NGC~3115}

The shape of the X--ray emission  follows closely that of the optical galaxy. 
The MECS isointensity contours appear elongated along the galaxy's major axis and
further indicate possible structure.  
Only a relatively short ROSAT PSPC
observation is available for this object.

The comparison of the radial profile of the total emission, obtained
from azimuthally averaged concentric annuli about the X--ray peak, with
the blank sky profile (Fig.~\ref{raw}) indicates an extension out to $\sim
9'$.  However, most of the emission outside $\sim 5'$ is due to the
source to the SW,
which is most likely unrelated to the galaxy, as shown by the photon
distribution obtained excluding a circle of $\sim 2\farcm3$ radius around the
SW source (see Fig.~\ref{raw}).

The comparison of the net profile with the PSF is shown in Fig.~\ref{net}.  The
NW quadrant (filled dots) is separated from the other three quadrants since it
appears less extended than the others and indeed it is consistent with the
radial distribution expected from a point source.  The other three quadrants
are instead significantly more extended and inconsistent with the PSF
distribution.  This could be the indication of a complex source composed of a
point-source embedded in a more extended and somewhat asymmetric
emission.  

For the spectral analysis however we cannot distinguish the two components and
we have therefore used a circle of radius $4'$ as suggested by Fig.~\ref{raw}.

\subsection{NGC~3379}

The X--ray contours indicate a complex source.  Most of the emission is 
associated with the target galaxy itself, but some emission is present in
coincidence with NGC~3384, in the region between NGC~3379 and NGC~3389, and to
the West of NGC~3379.   This latter source does not coincide with known
optical counterparts, but a point-like X--ray source is visible in the ROSAT
HRI image of this field. 
We have therefore excluded this source from the
spectral analysis of NGC~3379. 

Fig.~\ref{raw} shows the radial distribution of the total emission from
NGC~3379.  The azimuthally averaged profile (the interloper to the NW
is masked out with a 2$'$ radius circle) indicates emission out to a
radius \> 8$'$.  However, the emission outside $\sim 4'$ is due to the
extension to the S-SE.  In particular, the comparison between the net
emission in the Northern and Southern halves with the PSF indicate that
the central source is consistent with the photon distribution of a
point source.  The extension prominent in the SE quadrant is most likely due to
the presence of intergalactic material, but is too faint to be studied
with the BeppoSAX data. Although in a different energy band, the ROSAT
HRI data also suggest only a nuclear point source (cf. Roberts and
Warwick 2000).  We have therefore used an extraction radius of
4$'$ for the spectral analysis, to maximize the emission from the central
source and minimize that from the ``diffuse" component. 

\subsection{NGC~4125}

The MECS contour plot in Fig.~\ref{n4125-map} suggests a possible
flattening and an extent towards the south out to a radius of $\sim
5'$, and additional low surface brightness emission is visible outside
of the optical image, mostly to the W and NW.   LECS data in the same
hard band as MECS also suggest the same flattened source,  but in the
broader band shown in Fig.~\ref{n4125-map}, that includes softer
photons, the source appears rounder, most likely due to the larger
PSF.  The PSPC image (Fabbiano \& Schweizer 1995) also shows an equivalent
slight distortion of the galactic emission in the EW direction and
similar low surface brightness emission at similar distances from the
X--ray center.

The azimuthally averaged radial profile of the total emission indicates a sharp
decline out to r$\sim 3'$, outside of which the surface brightness flattens
and  matches reasonably well the shape of the blank sky fields
(Fig.~\ref{raw}).  
The LECS data instead show a slowly decreasing radial
profile of the total emission out to a radius of $\sim 6'$, outside of
which the profile is consistent with that from the blank sky
rescaled for the exposure times only.

We have attempted to determine whether the source is extended or
compatible with the \bsax PSF, and whether we can measure the distortion
apparent in the contour plot.  In spite of the poor statistical
significance of the observation, the radial profiles shown in
Fig.~\ref{net} for the full 2-9 keV energy range suggests that
there is an excess over the PSF at radii larger than 4$'$.  This is
however due to an excess in the NW quadrant, in the $\sim 4'-10'$ radii
annulus, excess that we detect both at softer (2-3 keV) and harder (3-7
keV) energies. 

For the spectral analysis, we have therefore considered a $4'$
radius circle for the extraction region in the MECS, that would exclude
the emission in the NW quadrant, and would only consider the emission
consistent with the PSF.  Since the LECS data indicate source counts out  to
$6'$ radius, we have used this larger extraction region for this
instrument.

\section{ASCA data}
\label{asca}

We have used the ``screened" data produced by the REV2 version of the
processing. 
Both sets of focal plane instruments were used (see Table~\ref{log}).
GIS data were obtained in PH mode, and SIS data in BRIGHT mode.
Full references to the ASCA instruments and observing modes can be found
in the on-line documents at {\tt http://legacy.gsfc.nasa.gov/docs/asca}.  
Since calibration is poor for
energies below 0.5 keV (SIS) and 0.8 keV (GIS), and above 5 keV for SIS,
we have restricted the analysis to energies between 0.5-5 keV and 0.8-10
keV for SIS and GIS respectively. 

Each galaxy is discussed briefly below, and the 
results from the spectral analysis are given in
\S~\ref{spectral} and tabulated in Table~\ref{spec}.

\begin{figure*}
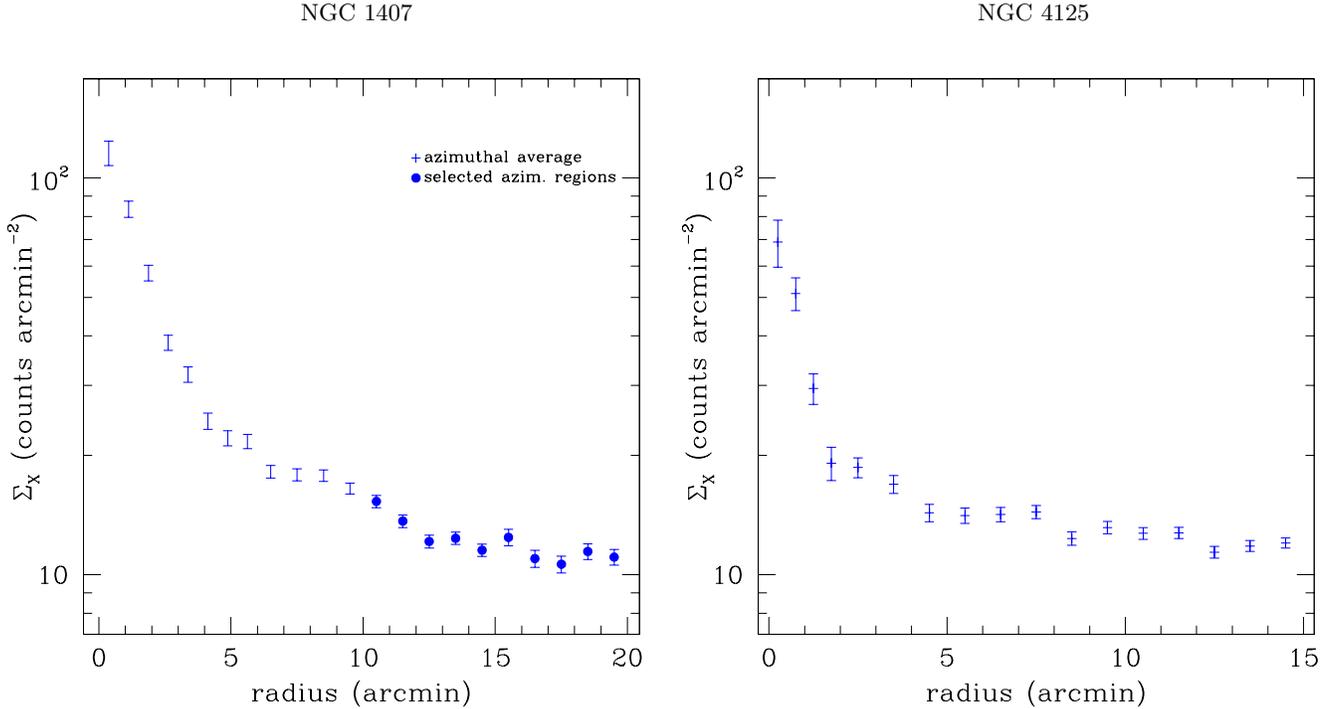

\unitlength1.0cm
\begin{picture}(18,10.0)
\thicklines
\put(0,0.0){
\begin{picture}(18,9.0)
\resizebox{18cm}{!}{
\psfig{figure=ms10097.f9a,width=16cm,clip=}
\psfig{figure=ms10097.f9b,width=16cm,clip=}
}
\end{picture}}
\put(4.0,9.2){
NGC~1407
}
\put(13,9.2){
NGC~4125}
\end{picture}
\caption[]{Radial distribution of the total emission observed with GIS
in the 0.8-10 keV energy range.  LEFT: data for NGC~1407. In order to avoid
regions outside the field of view, only angles -90\degree to 180\degree 
are averaged together at radii $10'<r<15'$, and angles -30\degree to
100\degree outside of r$=15'$ (angles are counterclockwise from N).
RIGHT: data for NGC~4125. The data are azimuthally averaged. }
\label{gis-prof}
\end{figure*}

\subsection{NGC~1407}
Both Einstein and ROSAT observations of this galaxy show the presence
of additional, possibly unrelated, fainter sources around the galaxy,
which are also visible in the ASCA images (Fig.~\ref{n1407-map}).
Moreover this galaxy is located within a group, so low surface
brightness emission from intergalactic gas is expected.  The radial
photon distribution indicates a slowly decreasing surface brightness
emission almost out to the edge of the field (the emission flattens 
only towards the NE edge off the field,
see Fig.~\ref{gis-prof}).  We have carefully considered which is the
best region to study the emission from this source.  In the GIS we have
defined a circle of 6$'$ radius, that corresponds to a flattening of
the surface brightness profile before it further decreases,
and is also the recommended size for inclusion of all the
flux from a strong point source.  
With this choice, we can retain a high signal-to-noise
and obtain the background locally in an adjacent region that should
include the contribution from the group, which we can subtract from the
galaxy's emission.  We have also considered a ``field" background by choosing 
a small region in the NE corner of the detector, where the profile shows that the 
surface brightness  is constant with radius.  This background was also used 
to study the emission from the region outside the galaxy ($i.e.$ the local
background considered above).

For both SIS, we have used a radius of $\sim 3\farcm3$, the maximum
that is allowed by the  offaxis position of the source in the detector,
that avoids including areas outside of the field of view in the source
region.  As for the GIS, the background is estimated locally around the
source, in a not-concentric annulus as large as it was compatible with
the detector size.

While it is true that the `unrelated' sources are included in the
background regions, their contribution should not be in excess of
$\sim$ 5\%, so that they will not significantly affect the background
characteristics.

\subsection{NGC~4125}

As shown by the profile in Fig.~\ref{gis-prof}, source counts are
visible out to a radius of $\sim 4'$, then the profile appears to
flatten, although a second plateau is visible outside $8'$. 
The source is relatively centered on the GIS field of view, but
not in the SIS.  We have therefore used a $4'$ radius circle for the
GIS, and chosen the background from a concentric annulus around the
source.  In the SIS, we have used the maximum radius, of $2\farcm 2$, to
avoid regions outside the CCD field of view, and a local background
from two regions close to the source.

\begin{table*}
\caption[]{Spectral Results }
\label{spec}
\bigskip

\begin{flushleft}
\begin{tabular}{lrrrrrrr} \hline
\noalign{\smallskip}
Model  & N$_H$& kT$_1$/$\Gamma$&
90\% conf.& kT$_2$& 90\% conf.& \chisq$_\nu$ &D.o.F.\\
&\multicolumn{1}{c}{(cm$^{-2}$)} &(keV) &\multicolumn{1}{c}{interval} 
& (keV) &\multicolumn{1}{c}{interval} \\
\hline
\hline
\noalign{\smallskip}
\multicolumn{8}{c}{NGC~1407: Combined GIS+SIS}\\
\multicolumn{8}{c}{
GIS2: 649$\pm$43 counts; GIS3: 852$\pm$48 counts; r=6$'$;}\\
\multicolumn{8}{c}{SIS0: 1518$\pm$53 counts; SIS1:1301$\pm$47 
counts; r=$3\farcm3$}\\ 
\hline
\noalign{\smallskip}
1 R & 5.5\e20 & 1.05 & &&&2.8 &133 \\
1 R $^\dagger$ & 5.5\e20 & 0.96 && && 1.7 &132 \\
1 R; 1 B & 5.5\e20&0.85&0.83-0.87&3.6 &2.7 -5.5 &1.0 &131\\
\hline
\hline
\multicolumn{8}{c}{NGC~1553--Combined MECS + LECS from both observations} \\
\multicolumn{8}{c}{MECS 305$\pm$28 and 207$\pm$25; LECS 209$\pm$21; r= 6$'$}\\
\hline
1 R & 2\e20 &3.2 & &&& 1.9 & 33 \\
1 R $^\ddagger$ &  2\e20 &3.4 &&&& 1.9 & 32 \\
1 R; 1 B &2\e20 &0.26&0.2-0.4 &4.8&3-8  & 1.0 & 31\\
\hline
\hline
\multicolumn{8}{c}{NGC~3115--Combined MECS + LECS from both
observations} \\
\multicolumn{8}{c}{MECS 391$\pm$28; LECS 95$\pm$15; r=6$'$}\\
\hline
1 R &4.5\e20 &9.5& 5-49  & &&0.8&21 \\
1 P & 4.5\e20 & 1.58 & 1.28-1.88& && 0.8 & 21 \\
\hline
\hline
\multicolumn{8}{c}{NGC~3379--Combined MECS + LECS} \\
\multicolumn{8}{c}{MECS 369$\pm$31;  LECS   100$\pm$16;  r=4$'$}\\
\hline
1 R & 3.8\e20& 6.9 & 4-24&&&1.5&15\\
1 P  &3.8\e20 & 1.8 & 1.4-2.1  &&& 1.3& 15 \\
\hline
\hline
\multicolumn{8}{c}{NGC~4125 --  Combined MECS + LECS} \\
\multicolumn{8}{c}{MECS 103$\pm$25 r=4$'$; LECS 136$\pm$18;  r=6$'$ }\\
\hline
1 R &2\e20&0.74&&&&2  &10 \\
1 R ; 1 B; &2\e20&0.33&0.25-0.5 &3.9&2-12 &0.26&8 \\
\hline
\multicolumn{8}{c}{NGC~4125 -- Combined GIS+SIS} \\
\multicolumn{8}{c}{%
GIS2: 136$\pm$19 counts; GIS3: 159$\pm$21 counts; r=4$'$;}\\
\multicolumn{8}{c}{SIS0:  385$\pm$22 counts; SIS1: 237$\pm$17
counts; r=$2\farcm2$}\\
\hline
1 R &2\e20&0.77&&&&2.7&59 \\
1 R; 1 B; &2\e20&0.35 & $^\star$& 1.6& $^\star$& 1.3& 47 \\
1 R; 1 B; &2\e20&0.63 & $^\star$& 2.4 &$^\star$&1.3 & 47\\
\hline\hline

\end{tabular}

\bigskip
Note:  The models assumed are: $raymond$ (R) with abundances fixed at
100\% cosmic values unless otherwise indicated; bremsstrahlung (B);
power law (P).  The 90\% confidence interval for
temperature(s) or photon index $\Gamma$  are for 1 interesting
parameter ($i.e$ they are derived from the \chisqmin+2.7 level) and are
calculated only for \chisqmin $\sim$ 1.  The N$_H$ is not well defined
by these data.  We have fixed it at the line-of-sight galactic value.
\\
$^\dagger$ The best fit gives abundances of 10\% cosmic.  $^\ddagger$
The best fit gives abundances of 0.1\% cosmic.
$^\star$  We find two minima of almost equal significance, well 
outside the 68\% error contours.  We have not derived 
errors for these parameters. 

\end{flushleft}
\end{table*}

\section{Spectral results}
\label{spectral}

We report in Table~\ref{spec} the results 
obtained from the spectral analysis.
The spectral data derived in the
regions discussed in previous sections have been binned to a coarser
energy resolution, to improve on the statistical significance in each
bin.  We have typically obtained bins in which the net counts are at
least 3$\sigma$ above the background (2$\sigma$ for the LECS data of
NGC~3115 and 2.5 $\sigma$ for the MECS data of NGC~4125).  
For each observation, we have
fitted simultaneously all instruments, fixing the spectral parameters
equal for all of them, but we have left the relative normalization
as a free parameter, which compensates for different extraction
regions in different instruments and cross correlation
uncertainties.  We have first used a single model
with low energy absorption (mostly a thin plasma model [$raymond$ in
XSPEC], with cosmic abundances fixed at 100\%, or a bremsstrahlung).
Since in most cases the fit gave a high $\chi^2_{\nu}$ or significant
residuals in the model-data comparison, we have tried both to relax
the constraint on abundances and we have added a second component to
the model.  Variable abundance fits produced only a partial
improvement in $\chi^2_{\nu}$, and gave extremely low, subsolar
values.  The addition of the second
spectral component gave acceptable $\chi^2_{\nu}$ values close to 1,
even in cases where the statistical significance of the data is poor.
A further attempt to lower the $\chi^2_{\nu}$ value with a variable
abundance model for the soft component does not lead to improvements
(see NGC~1407).  Since the statistical significance of the LECS data
is poor, and the SIS data also did not give significant constraints on
the low energy absorption value, we have fixed the low energy
absorption at the line-of-sight galactic value (Dickey \&  Lockman
1990).

Individual galaxies are discussed below. 
  
{\bf NGC~1407}:
The results reported in the table refer to the combined SIS and GIS fits 
to the ``galaxy" regions with a local background.  We briefly discuss
below also the results to the GIS data obtained in the outside region
described in \S~\ref{asca}.

A single temperature model is not an acceptable fit either to  the
single data sets or the combined data (see Table~\ref{spec}). 
When the assumption of fixed abundances is relaxed,
the improvement in the \chisqmin\ is significant, but not sufficient to 
bring it to acceptable values, and the  distribution of the residuals
confirms that this would be a poor fit.
We have then added both a second
$raymond$ model with 100\% cosmic abundances, and a thermal
bremsstrahlung: we found that a component at kT $\sim$ 0.9 keV, and
one at kT $\sim$ 3-5 keV give a good fit to the data, for either choice
of model for the second component.  
  
The reduced \chisqmin indicates that the model is adequate, and
therefore there should be no reason to supply more free parameters to
the fit.   We have nonetheless also tried the same models, but
letting the abundance parameter free to vary, to check our results
against those of other authors, who report less than solar best fit
abundances (Awaki et al. 1994, Matsumoto et al. 1997).  The 
improvement in \chisqmin is actually none, and the abundance parameter
cannot be constrained (only $>10$\%).  

We have also substituted the $raymond$ with the MEKAL model, for the
low temperature component, since this latter should better model the Fe
complex at $<$1 keV.  The results are qualitatively consistent with
those reported in Table~\ref{spec}.  We find that the best fit values are
somewhat different (lower temperatures, as reported by several authors) 
with equivalent goodness of fit  ($\chi^2_\nu \sim 0.94$).  

Outside of  $6'$ the statistical significance of the data decreases
considerably, and the effects of the photons scattered at larger radii
because of the Point Spread Function of the ASCA GIS have not been
properly modeled (see ASCA on-line documents).  
However, a fit to the data
indicates that the spectrum can be parameterized by a $raymond$
spectrum with kT$\sim 1.4$ keV, consistent with intergalactic gas in a
group.

{\bf NGC~1553}: 
Since the two halves of the observations were obtained with different sets of instruments
(see \S~\ref{datasax}), we had to consider three separate sets of data:  MECS data for
observation 1 (3 instruments operating), MECS data for observation 2 (2
instruments operating) and LECS data for both observations.  

The 2-temperature fit improves significantly the values of the 
minimum $\chi^2_\nu$ relative to the single temperature model, 
and indicates temperatures of 
kT$_1$ $\sim$0.3 and kT$_2$ $\sim$5 keV. While the need for a
two-component model is evident,  the requirement of a thin plasma 
model for the low energy component is not stringent: the $\chi^2_{min}$ values 
are comparable when using a $raymond$ (Table~\ref{spec})
or a bremsstrahlung model (best fit temperatures kT$_1 \sim 0.5$ and
kT$_2 \sim 11$), although some positive residuals can be seen at energies
below 1 keV in the latter case.

{\bf NGC~3115 and NGC~3379}: A one-temperature component is adequate to fit the
data of either galaxies.  Equivalent goodness of fit is obtained from a $raymond$, a
bremsstrahlung or a power law model.  

{\bf NGC~4125}:
The spectral fit both from  the \bsax and from the ASCA data indicates once
again that a single temperature model cannot represent the data.

In the 2-temperature fit we find that the soft component must
be parameterized by a thin plasma code ($raymond$ in this case):
assuming a Bremsstrahlung model for the soft component
gives a higher minimum $\chi^2_\nu$ =2, 
and leaves significant positive residuals at energies below 1
keV.  Other models we have attempted for the low energy component, 
like a black-body spectrum, give
equally bad distribution of the residuals.

\subsection{Fluxes and luminosities of the different components}

Table~\ref{fluxes} summarizes the best fit fluxes and luminosities of the
galaxies discussed here.  Total band values as well as the contribution of each
component are given in several bands, for easy comparison with previous works.
We have converted the MECS or the GIS count rates to fluxes using the best fit
spectral parameters reported in Table~\ref{spec}, and the distances in
Table~\ref{gen} to convert these in luminosities. 

For NGC~4125, we have used the LECS instead of the MECS data, since
the former give a flux consistent with the GIS flux.  MECS data give a
significantly lower flux.  Given the peculiarly shaped extraction region
(see \S~\ref{datasax}), the NGC~1553 flux is derived from the counts
listed in Table~\ref{spec} rescaled to the total expected number,
assuming an azimuthally symmetric source (factor of $\sim 1.4$).

\begin{table*}
\caption[] {Fluxes and Luminosities in different bands for the deferent 
spectral component. 
 }
\begin{flushleft}
\begin{tabular}{ l l  l l  l  l| l  l l  l }
\noalign{\smallskip}
\hline
\noalign{\smallskip}
Name &&\multicolumn{4}{c|}{Intrinsic Flux} &\multicolumn{4}{c}{Intrinsic 
Luminosity} \\
&&0.1-2 keV&0.2-4 keV &0.5-5 keV &2-10 keV&0.1-2 keV&0.2-4 keV&0.5-5 keV
&2-10 keV \\
\noalign{\smallskip}
\hline
\noalign{\smallskip}
\hline
\noalign{\smallskip}
NGC1407 &T&1.8$\times 10^{-12}$&1.9$\times 10^{-12}$&1.7$\times 10^{-12}$&
6$\times 10^{-13}$&2.6$\times 10^{41}$&2.8$\times 10^{41}$&2.5$\times 10^{40}$&
8.9$\times 10^{40}$\\
&s&1$\times 10^{-12}$&9$\times 10^{-13}$&8$\times 10^{-13}$&5$\times 
10^{-14}$&1.4$\times 10^{41}$&1.3$\times 10^{41}$&1.2$\times 10^{41}$&
7$\times 10^{39}$\\
&h&7.5$\times 10^{-13}$&1$\times 10^{-12}$&9$\times 10^{-13}$&
6$\times 10^{-13}$&1.1$\times 10^{41}$&1.4$\times 10^{41}$&1.3$\times 10^{41}$&
8$\times 10^{40}$\\

NGC1553 &T&2.0$\times 10^{-12}$&2.0$\times 10^{-12}$&1.7$\times 10^{-12}$&
7$\times 10^{-13}$&1.1$\times 10^{41}$&1.1$\times 10^{41}$&9.7$\times 10^{40}$&
4$\times 10^{40}$\\
&s&1.3$\times 10^{-12}$&1.2$\times 10^{-12}$&8$\times 10^{-13}$&
7$\times 10^{-16}$&7$\times 10^{40}$&7$\times 10^{40}$&5$\times
10^{40}$&4$\times 10^{37}$\\
&h&7$\times 10^{-13}$&1$\times 10^{-12}$&9$\times 10^{-13}$&
7$\times 10^{-13}$&4$\times 10^{40}$&5$\times 10^{40}$&5$\times 10^{40}$&
4$\times 10^{40}$\\

NGC3115&P&3$\times 10^{-13}$&5$\times 10^{-13}$&4$\times 10^{-13}$&
5$\times 10^{-13}$&5$\times 10^{39}$&6$\times 10^{39}$&6$\times 10^{39}$&
7$\times 10^{39}$\\
 
NGC3379&P&3$\times 10^{-13}$&4$\times 10^{-13}$&3$\times 10^{-13}$&
3$\times 10^{-13}$&6$\times 10^{39}$&7$\times 10^{39}$&7$\times 10^{39}$&
6$\times 10^{39}$\\

NGC4125&T&7$\times 10^{-13}$&7$\times 10^{-13}$&6$\times 10^{-13}$&
1$\times 10^{-13}$&1.2$\times 10^{41}$&1.2$\times 10^{41}$&1$\times
10^{41}$& 2.4$\times 10^{40}$\\
&s&5$\times 10^{-13}$&5$\times 10^{-13}$&4$\times 10^{-13}$&1
$\times 10^{-15}$&9$\times 10^{40}$&8$\times 10^{40}$&7$\times 10^{40}$&
2$\times 10^{38}$\\
&h&2$\times 10^{-13}$&2$\times 10^{-13}$&2$\times 10^{-13}$&
1$\times 10^{-13}$&3$\times 10^{40}$&4$\times 10^{40}$&4$\times 10^{40}$&
2$\times 10^{40}$\\

\noalign{\smallskip}
\hline
\end{tabular} 
\end{flushleft}
\medskip

Total (T), soft (s), hard (h) components from the two-component fits from
Tables~\ref{spec} are given.  Power law fits are used to compute the flux of
NGC~3115 and NGC~3379.  All quantities have been corrected for the Galactic
line of sight value of N$_H$.  Values are from MECS data (GIS for NGC~1407) 
except for NGC4125, where the LECS values are used, since they are comparable
to the GIS values.   NGC~1553 values have been rescaled by a factor of 1.44
to compensate for the area that has been masked out (see text). 

\label{fluxes}
\end{table*}

\section{Summary of results}

The spatial analysis on the MECS data indicates that the distribution of
the hard emission in the four galaxies observed with \bsax\ is different
in different cases:  NGC~1553 is clearly extended, while NGC~3379 is
most likely dominated by a nuclear source.  A significant contribution
from a point source cannot be excluded by the data of NGC~3115, although
a more extended component is also observed.  The data of NGC~4125 are
not conclusive.

For three sources (NGC~1407, NGC~1553 and NGC~4125) the data require
a two component spectrum, with a soft and a hard component,  although
neither can be  determined with high accuracy.  
The soft component is well parameterized by a thin plasma spectrum,
with kT$\sim 0.3- 0.9$ keV, and in all three cases its contribution is
comparable or higher than that of the hard component in the soft
energy band.

At hard energies, the extended nature of the emission in NGC~1553
ensures that a nuclear source, if present, does not contribute
significantly to the observed luminosity.
Both NGC~3115 and NGC~3379 data instead could be parameterized by a power law
spectrum, consistent with a significant contribution from a nuclear source.  

An intriguing result is provided by the lack of a very soft
component in NGC~3115 and NGC~3379, commonly found in the lowest $\rm
L_x/L_b$ galaxies  (see discussion in \S~\ref{Intro}).  
While the poor statistics does
not allow us to speculate on whether this is a ``significant" result,
in the sense of singling out these two galaxies for their lack of a
very soft component, the existence of a nuclear component probably
further complicates matters.  In fact, the soft band ROSAT data of
either galaxies do not show a significant contribution from extended
components: the galaxies appear rather compact both in the PSPC image
of NGC 3115 and in the HRI image of NGC~3379 (cf. Robert \& Warwick 2000), 
suggesting that the
nuclear source dominates at all energies.  We have estimated the
contribution from a soft component allowed by the present
data.  As discussed before, one component is adequate, and in fact the
addition of a second component to either set of data does not give any
additional information: the second component either mimics the
parameters of the first component or has a zero normalization.  To
estimate a possible contribution from a soft component, we have
therefore fixed the power law parameters at the best fit values and
added a $raymond$ with kT = 0.3 keV.  We find that in either cases a
soft component is allowed, and would have an unabsorbed flux
$f\rm _x(0.1-2 keV) \sim 1-3 \times 10^{-13}$ \ergcms (NGC~3115 and NGC~3379
respectively).  We will discuss this further in \S~\ref{discuss}.

\section{Comparison with previous spectral results}

The ASCA data of NGC~1407 and NGC~4125 have also been analyzed by 
Buote \& Fabian (1998) and by Matsushita et al. (2000).   In the comparison
between this and the other two works, we cannot expect equal results,
since the spectral
analysis is based on different assumptions in all three cases:  Buote \&
Fabian assume a different spectral code for the soft emission (MEKAL)
with variable abundances; Matsushita et al.  assume a fixed 10 keV
temperature for the hard component, and variable abundances for the
soft component.  However, in spite of the different assumptions,
we find good agreement in the best fit soft temperature 
of NGC~1407, and to within a factor of 2 for the 
fluxes and luminosities of the different components. 
For NGC~4125, we cannot properly determine the
best fit parameters from the ASCA data, and the Matsushita et al. low
temperature component is intermediate between our two best fits. 

The ROSAT data of these galaxies have been discussed 
by Irwin and Sarazin (1998).  However, these authors do not perform a   
spectral analysis of the data, and the hard and soft luminosities they
give do not come from two spectral components, but are defined as the  
luminosity in two narrow bands within the ROSAT band.   Therefore they
are not directly comparable with the ASCA and \bsax\ values. 

In the PSPC data of NGC~4125, Fabbiano \& Schweizer (1995) find two
spectral components, at $\sim 0.2-0.3 $ keV, and $\sim 0.5-0.6 $ keV
and suggest that the harder one is evidence of hot gas.   In the \bsax\
data we do not find the harder component of Fabbiano \& Schweizer.
However, we notice that the ASCA data do show two minima of equal
significance at $\sim 0.3 $ and $\sim 0.6 $ keV (Table~\ref{spec}).
A similar result was found in the PSPC data of NGC~1553 by Trinchieri
et al (1997), where one of the two components could have 
an intermediate $\sim 1$ keV temperature, when the low 
energy absorption is at the Galactic line-of-sight value.

\section{PDS detection}

A $\ge 4\sigma$ detection is obtained with the PDS during
both observations of the NGC~1553 field. 
The PDS detector is a non-imaging instrument with an
hexagonal FoV with FWHM $\sim 75'$ (Frontera et al. 1997 and
references therein). Given the very large field of
view of this instrument and the lack of spatial resolution, the
identification of the PDS source is not straightforward.  

We have used the PHA files
provided by the SDC to study the origin of this emission, and in
particular to understand whether the source detected with the PDS could
be related to the galaxy.

We have first tried to fit the PDS data together with MECS and LECS.  
The best fit model found for the LECS+MECS combination could fit the
PDS data, but would require a
normalization factor of $\sim 70$ for the PDS data, well outside the
expected value of $\sim 1$ ($\sim 0.8-0.9$ for MECS-PDS and $\sim
0.8-1.2$ for LECS-PDS).  This indicates a very
large excess over the emission expected from the galaxy at higher
energies, and argues against an association with it.  

We have therefore searched for other plausible candidates in the  PDS
FoV. As shown by the MECS data, there are virtually no other
sources in the field, also when we consider hard energies only (above
4  and 6 keV to account for heavily absorbed sources), and a similar
search in the 2 degree PSPC field centered on NGC~1553
also suggests that all of the ROSAT
sources in the area would have been in the MECS FoV, indicating that
they are either very soft or too faint to be detected by \bsax.  It is
therefore highly unlikely that any of these is the counterpart of the
PDS source.  On the other hand, it is also unlikely that the
combination of all these sources causes the PDS detection: NGC~1553 is
by far the brightest at high energies, and the discrepancy in flux is
too large to be accounted by the sum of a few much fainter sources.

We have searched for neighbouring sources in an area of 1.5  degree radius
around NGC~1553 in X--ray catalogs.  We have found entries only in the
catalogs from ROSAT and $Einstein$ data.   
In the RASS source list we found 8 
sources, two of which are contained in the MECS
FoV (one in fact is NGC~1553 itself).  The brightest one, at $\sim$
1\degree NE, coincides with NGC~1566,  which is also detected in the
$Einstein$ Slew Survey.  This is the only possible source of emission
since it is a Seyfert galaxy, with a variable soft X--ray flux measured
by $Einstein$ up to f$_{0.2-4.0 } \sim 2.2 \times 10^{-11}$ erg
cm$^{-2}$ s$^{-1}$ (Fabbiano, Kim and Trinchieri 1992).  

\bsax\ observations of NGC~1566 do not exist, and we cannot therefore
unambiguously establish that the PDS detection is indeed due to the
emission of NGC~1566. One of the main problems is that all observations
available from previous X--ray missions ($Einstein$ and ROSAT)
are in a much softer band than
the PDS, so that comparison of the flux in a common band has to rely on
the spectral parameters assumed, that have to be extrapolated over more
than a decade in energy.  Since the spectral capabilities of both
previous missions and the PDS instrument, with the low statistical
significance of the data, are poor, this comparison is very uncertain.
We have applied  different spectral models to the PDS data and derived 
band fluxes for comparison with  previous results:  a) the best fit
power law $\Gamma$=2.3 determined from the ROSAT data (Ehle et al.
1996) gives f$_{2-10}\sim 3\times 10^{-11}$ erg cm$^{-2}$ s$^{-1}$ 
and  f$_{0.2-4.0}  \sim 1   \times
10^{-10}$ erg cm$^{-2}$ s$^{-1}$ b) the
canonical power law slope  for Sey 1, $\Gamma$=1.7, gives
f$_{2-10} \sim 2 \times 10^{-11}$ erg cm$^{-2}$ s$^{-1}$
and f$_{0.2-4.0}  \sim 2\times
10^{-11}$ erg cm$^{-2}$ s$^{-1}$.

Given all of the uncertainties in the spectral parameters, the
extrapolation from the PDS to the softer energy band is consistent with
the identification of the PDS source with NGC~1566, in particular if 
the source was in a high flux state, since the above
estimates do not take into account the degradation
of the PDS response at the large off-axis position of the galaxy.

\section{What is the origin of the X--ray emission in early-type
galaxies?}
\label{discuss}

The presence of at least 2 components in the X-ray spectra 
of early type galaxies is now well established for all galaxies 
spanning the whole range of L$\rm _x/L_b$ ratio, although the origin of
these component might be different, as will be discussed later.

Even though the quality of the present data does not allow us to derive very
precise spectral parameters for the galaxies studied here, we can
discuss their properties in light of these results together with current
discussions in the literature. 

When the full spectroscopic range of $\sim 0.2 -10$ keV is taken into
account, the need of at least two components to the model fit becomes
quite stringent.  As already noticed often in the literature (cf. Buote
1999 and references therein), the present data stress the
importance of including the
contribution of the harder spectral component at soft energies, even
though it might not be possible to measure its strength or detailed
characteristics, to avoid 
a faulty interpretation of the data.

A thin plasma spectrum is most likely the best model 
to fit the low temperature component in these intermediate and low L$\rm _x/L_b$
objects. For NGC~1407 and NGC~4125 a simple thermal bremsstrahlung fails to
reproduce the data at energies below 1 keV, indicating either the need
of an additional component or of line emission.  
Lines are also required to model the low energy component of
X--ray bright objects, 
interpreted as due to the emission of hot gas, although the detailed
characteristics (mostly abundances), of the lines
have not been unambiguously determined at the present time. 
In the lower L$\rm _x/L_b$ galaxies, the presence of 
lines cannot be used to firmly identify the soft component with
emission from hot interstellar gas, since line emission below $\sim 1
$keV has also been found in the spectra of Galactic low mass binaries
(Her X-1 and 4U1626-67,  Oosterbroek et al.  1997; Owens et al.
1997).

In fact the presence of hot gas at the low end of the X--ray
luminosity distribution has been recently challenged by Irwin and
Sarazin (1998) who, based on the similarity between the X--ray colors
(in the ROSAT band) of low X--ray luminosity galaxies and the bulge of
M31, NGC~1291 and galactic low Mass X-ray Binaries (LMXB), have
suggested that the LMXB sources could account for the whole spectrum of
low $\rm L_{\rm X}/L_{\rm B}$ galaxies.  \bsax\ and ASCA data on the bulge
of M31 (Trinchieri et al.  1999), together with a new analysis of ROSAT
PSPC data (Irwin \& Bregman 1999),   have confirmed the need of a soft
component to model the 0.1-10 keV spectrum.  The spatial evidence from
high resolution X--ray images indicates that a major fraction of the
emission is resolved into individual sources, while a smaller percentage 
could be accounted for by either a population of lower luminosity sources or
by a truly diffuse component (the fraction attributed to each component
however is different in the $Einstein$ and ROSAT data, see Trinchieri
and Fabbiano 1991 and Primini et al. 1993). This 
argues against a significant contribution from an ISM in
the bulge of M31, and suggests that the soft emission detected could indeed be
attributed to the low mass binary population. 
However, whether LMXBs could account for $all$ of this component, and
whether the same 
applies to low luminosity early type galaxies as well is still a
very open question:  1) the relative contributions of the soft and hard
components  in M31 do not appear to be the same as those measured in
early type galaxies, at least based on the evidence given by \bsax\ and
ASCA (Trinchieri et al. 1999; however, see Irwin and Bregman 1999);
2)  Kim et al.  (1998) discuss the positive detection of
gas in NGC~1316, one of the low L$\rm _x/L_b$   early type galaxies,
which further reinforces the need of more than just binaries in these
objects; 3) Borozdin \& Priedhorsky (2000) argue that the spectra of the
diffuse emission and of the binary sources in M31 differ, suggesting that the
soft component is to be attributed all to the unresolved component and
not to the LMXB population. 

\begin{figure}
\psfig{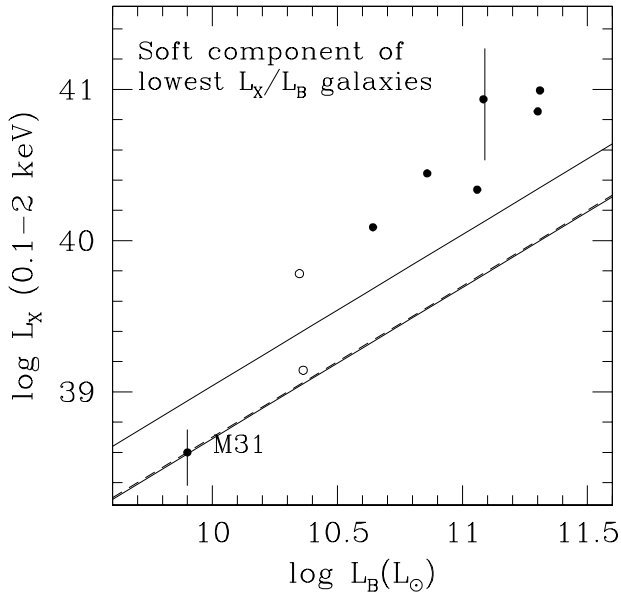}
\caption{\label{corr} Correlation between the optical (B-band) and
soft X--ray luminosity (0.1-2 keV band) in the lowest L$\rm _x/L_b$
early type galaxies (i.e., those belonging to group 1, as defined in
Table 1).  The solid lines represent the expected contribution from
LMXBs estimated by Irwin \& Bregman (1999) from an
$ASCA+ROSAT$ analysis of the bulge of M31; the 
lines shown include the extreme contributions allowed by different spectral
models that are considered equally good by the authors.  The dashed
line is normalized to the estimate of the emission of a very soft 
component in the bulge of M31 from $BeppoSAX$ data (Trinchieri et al. 1999).
Data are from this paper for NGC4125, Fabbiano et al. (1994) for
NGC~4382 and NGC~4365, Kim et 
al. (1996) for NGC~1316, Pellegrini (1994) for NGC~5866, and Matsushita 
et al. (2000) for NGC~4697. The open symbols indicate the estimates of the
soft component in NGC~3115 and NGC~3379 (see text). 
Errors on the flux are given for NGC~4125 and M31 and are estimated
from the 90\% uncertainties on the normalization of the soft spectral
component.  
}
\label{soft}
\end{figure}

Figure~\ref{soft} shows the comparison between the soft emission observed
in the bulge of M31 with \bsax, ASCA and ROSAT 
and that observed in the lowest L$\rm _x/L_b$   early type galaxies.
In all objects where the soft component has been measured 
there is excess emission over even the higher estimate of the expected
contribution from binary sources, suggesting that an additional
component should be present or that the binaries contribute a
higher proportion in the lowest L$\rm _x/L_b$   galaxies.  
Better quality data however, 
such as will be provided by $Chandra$ and XMM-Newton observations,
are needed to understand the nature of the soft emission in early type galaxies.

At the opposite end of the spectrum, the origin of the
emission that dominates at higher energies remains unclear.  The evolved stellar
population in early type galaxies should give rise to hard emission
proportional to the stellar content, similar to what is observed for
late type galaxies.  In addition, there could be
emission from a nuclear source, although in most cases a further
requirement that the source is heavily obscured is necessary when the 
high resolution data at soft energies do not show evidence of an unresolved
nuclear source.  If the
evolved stellar population is responsible, we expect the spectral
characteristics of the  hard
component to be similar to those of the 
LMXB in our Galaxy, in M31 and to the integrated emission
in late type galaxies, where
this component is thought to dominate.  We also expect its
luminosity to be almost linearly correlated to the optical luminosity, as in the
spiral galaxy sample (Fabbiano et al. 1992), 
indicating that a constant fraction of
stars has evolved in the compact binary systems.  Moreover, the
emission should be clearly extended, and follow the radial
distribution of the optical light.  Signatures of nuclear activity at
other wavelengths combined with no spatial extension and appropriate
spectral characteristics would instead point to a nuclear source in
the galaxy.

The existence of the hard component was first suggested on the basis of
very crude spectral data from the $Einstein$ IPC (Kim et al. 1992)
but was measured for
the first time with ASCA data in an increasingly large number of
galaxies (Matsushita et al.  1994; Matsumoto et al. 1997).
However, there is to date no firm determination of its spectral
characteristics: in most cases (and the
\bsax\ and ASCA data presented here are consistent with this),
either a power law or a thermal model can be used to parameterize
its spectrum equally well (see also Allen et al. 2000, 
Table~\ref{spec}).

The temperature measured for the high energy component (kT in the
range $\sim 3-10$ keV, see Table~\ref{spec}) is consistent
with the spectral characteristics of the LMXBs
in our Galaxy (van Paradijs 1998) and of the bulge and several
sources of M31 measured with \bsax (Trinchieri et al. 1999).  
For power law spectra, the  indexes reported  in Table~\ref{spec} appear 
to be consistent with the canonical slope of Seyfert galaxies
(see also Matsumoto et al. 1997, Buote \& Fabian, 1998).  This is not
consistent with the results for a sample of high X--ray
luminosity galaxies studied by Allen et al. (2000), whose hard
component appears to
have a flatter spectrum.  However, the flat component in the very high
luminosity objects is in general a small fraction of the emission in
the hard band and appears to dominate only at very high energies
(above 5 keV in the spectrum of M87, see Fig. 2 in Allen et al. 2000),

The spatial analysis of the \bsax\ data indicates that the high energy
emission of NGC~1553 is extended and that a point source and an extended 
component account for the emission in NGC~3115
(Fig~\ref{net}).  This argues against a dominant contribution from
the nucleus in these two objects (although in NGC~3115 the two
could contribute equally).  The \bsax\ data on a third object, NGC~3923
(Pellegrini 1999), also argue against nuclear emission. 
For the other two galaxies observed
with \bsax, NGC~3379 and NGC~4125, the spatial evidence is not as
conclusive: there is an indication of excess emission over the PSF, 
but mostly due to other, possibly unrelated, components.  
However, a nuclear contribution is likely to be important in NGC~3379,
since Roberts and Warwick (2000) report the detection of a single 
nuclear source in this galaxy from HRI data. 

We therefore can exclude a nuclear origin of the hard emission only for
NGC~1553 (and NGC~3923), for which a stellar origin of the emission is
most likely.  From the spatial analysis of NGC~3115 we expect a mixture
of nuclear plus stellar component in this galaxy.   A large
contribution is expected from the nucleus of NGC~3379. 

\begin{figure}
\psfig{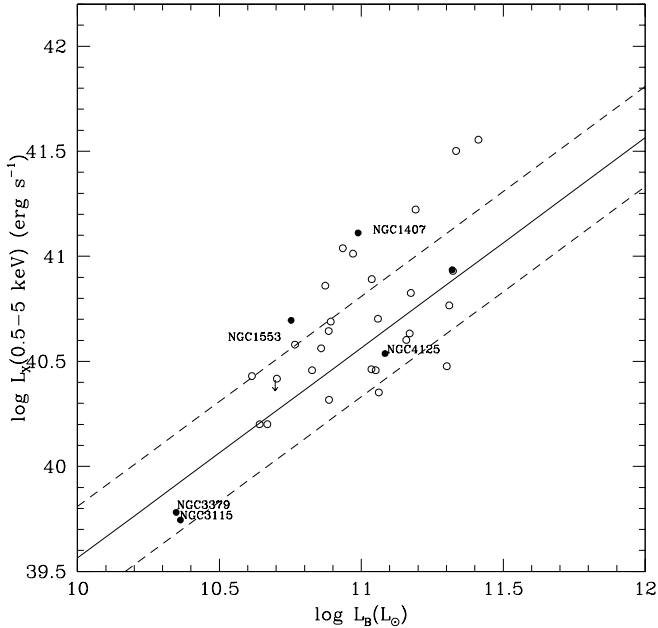}
\caption{\label{corr} Correlation between the optical (B-band)
luminosity and that of the hard component, with temperature $\ge$ 3
keV, for early type galaxies. The galaxies studied here are labeled
with their names.  Data for the other galaxies are from 
Buote and Fabian (1998), Kim et al. (1996), Matsumoto et
al.  (1997), Matsushita et al . (2000).  The nameless dot is NGC3923, also
studied with \bsax\ and also found extended in the hard band (Pellegrini
1999).  The solid line
represents the expected contribution from the binary component 
(Canizares et al. 1987); this estimate is coincident with that 
obtained from ASCA data by Matsumoto et al. (1997). The dashed lines
enclose the expected scatter/uncertainty for the binary contribution 
(Canizares et al. 1987).}
\end{figure}

We have tried to compare the characteristics of
the hard emission of these galaxies to those previously reported in the
literature, in an attempt to discriminate between 
binary and nuclear emission.  
In Fig.~\ref{corr} we present a plot similar to Fig. 7 of
Buote \& Fabian (1998), but we have included 
$only$ galaxies for which the hard component has a temperature higher than
$\sim$ 3 keV, so that we are sure we are not including the contribution
from the hot gas, or an hotter phase of
the ISM (see Buote \& Fabian 1998).

As shown by the figure, the scatter in the points is relatively large
and the present galaxies are in very good agreement with the other
early type galaxies from the literature.  All are consistent with what
might be expected from emission from the binary population (sketched
by the lines in the figure, derived by Canizares et al. (1987), based
on globular clusters and sources in the bulge of M31), as already
noticed by Buote \& Fabian (1998).  The extended nature of the emission
in NGC~1553 and NGC~3923 guarantees that these
estimates are indeed reasonable, as shown by their positions in the
plot.  NGC~1407 appears to be outside the range of the binary sources. 
However, since it hosts a compact, flat-nuclear source (Disney \& Wall
1977, Dressler \& Wilson 1985) and it is located at the center of a
small group, we cannot exclude additional  emission from either of
these components.  A similar reasoning can be applied to other galaxies
in the same part of the diagram.

Stellar kinematics for two of the present galaxies, NGC 3115 and NGC
3379, testify to the presence of a central mass concentration of
M$\sim 10^8-10^9$ M\sun (Kormendy et al. 1996; Emsellem 1999,
Gebhardt et al. 1996; 2000; Magorrian et al. 1998).  NGC~3379 has
been classified as a LINER 2 in the spectroscopic survey to search
for nuclear activity conducted by Ho et al. (1997). A contribution
from the nuclei could therefore be expected in these two objects.
However, their hard components are within the range expected for the
binaries contribution (Fig. 11), and their luminosities are well
below what has been observed in other galaxies who host a similar or
even lower mass nuclear black holes, $e.g$ NGC~4258 (Fiore et
al. 2000), with an X--ray luminosity L$\rm _x(2-10) \sim 10^{41}$ \ergs
and a black hole mass of $\sim 4 \times 10^7$ M\sun, or the Sombrero
galaxy, with L$\rm _x(0.1-2.4) \sim 3.4 \times 10^{40}$ \ergs (Fabbiano \&
Juda 1997).  

If we assume a substantial contribution from
the nucleus, then the contributions from binaries alone is lower
(from the spatial analysis, a factor of about 2 for
NGC~3115, most likely more for NGC~3379), and could fall below the      
estimates shown in Fig.~\ref{corr} for the binary contribution. 
However, given the relatively
large range spanned by the points, the poor knowledge of the
properties of this component in early type galaxies and the 
lack of points at the same low optical
magnitude of these galaxies, we cannot realistically speculate 
on whether this would make the high energy emission from these two
galaxies peculiar relative to the other objects. 

We further notice that the (unknown) intrinsic scatter in the plot
could be enhanced both by the limited quality of the data available and
by the limited information available on the detailed properties of
these sources.  Different assumptions on the spectral models,  
the uncertainties even within a single
set of spectral model and the presence of
several components (nuclei, binaries, group emission), which cannot be
discarded both in different objects and within the same object with the
present data, could 
introduce factors of a few in the estimates of the source luminosities
(as already remarked in \S 7).  

We can therefore only  conclude that on average we expect that a large 
fraction of the hard emission
in early type galaxies originates from the evolved stellar population,
as seen in later types.   Once again, the spatial and spectral
capabilities of $Chandra$ and XMM-Newton are required to properly
measure the characteristics of this component and therefore 
better understand its origin.

\begin{acknowledgements}
This work was partially supported by the Italian Space Agency.  
GT thanks Prof. Tr\"umper and the MPE staff for their hospitality and
support while part of this work was done. 
This research has made use of the NASA/IPAC Extragalactic Database
(NED)   which is operated by the Jet Propulsion Laboratory, California
Institute   of Technology, under contract with the National
Aeronautics and Space Administration.

\end{acknowledgements}


\begin{thebibliography}{}


\bibitem{} Allen, S.W., Di Matteo, T. \& Fabian, A.C. 2000, MNRAS 311 493

\bibitem{} Awaki, H., Mushotzky, R., Tsuru, T., Fabian, A. C., Fukazawa, Y.,
    Loewenstein, M., Makishima, K., Matsumoto, H., Matsushita, K.,
	  Mihara, T., Ohashi, T., Ricker, G. R., Serlemitsos, P. J., Tsusaka, Y.,
		 Yamazaki, T. 1994, PASJ 46L, 65.

\bibitem{} Boella, G., Butler, R. C., Perola, G. C., Piro, L., Scarsi,
           L., Bleeker, J. A. M. 1997a, AAS 122 299

\bibitem{} Boella,G., Chiappetti, L., Conti, G., Cusumano, G., Del Sordo, S.,
La Rosa, G., Maccarone, M. C., Mineo, T., Molendi, S., Re, S., Sacco, B.,
 Tripiciano, M.  1997b AAS 122 327
 
\bibitem{} Borozdin, K.N. \& Priedhorsky, W.C. 2000 astroph

\bibitem{}  Buote, D.A.  1999 MNRAS 309, 685

\bibitem{}  Buote, D.A.  \& Fabian A.C. 1998 MNRAS 296 977

\bibitem{} Canizares, C.R., Fabbiano, G., Trinchieri, G. 1987 ApJ 312,
503

\bibitem{} Dickey, J.M., Lockman, F.J. 1990, ARAA 28, 215 

\bibitem{} Disney, M. J., Wall, J. V 1977 MNRAS 179, 235

\bibitem{} Dressel, L. L., Wilson, A. S 1985 ApJ 291 668


\bibitem{} Ehle, M. Beck, R., Haynes, R.F., Vogler, A., Pietsch, W.,
Elmouttie, M., Ryder, S. 1996, AA 306, 73.

\bibitem{} Emsellem E. 1999 astro-ph/9910006

\bibitem{} Fabbiano, G. Kim, D-W. and Trinchieri, G. 1994 ApJ 429, 94 

\bibitem{} Fabbiano, G., Kim, D-W., and Trinchieri, G. 1992, ApJS 80, 645

\bibitem{} Fabbiano, G., Juda,  1997 ApJ 476, 666 

\bibitem{} Fabbiano, G., Schweizer F. 1995 ApJ 447 572

\bibitem{} Fiore. F., Giommi, P., Vignali, C., et al. 2000 MNRAS
submitted

\bibitem{} Frontera, F., Costa, E., Dal Fiume, D., Feroci, M.,
 Nicastro, L., Orlandini, M., Palazzi, E., Zavattini, G. 1997 AAs 122 357

\bibitem{}  Gebhardt, K., Richstone, D., Ajhar, E. A.,
   Lauer, T. R., Byun, Y.-I., Kormendy, J., Dressler, A., Faber, S. M.,
      Grillmair, C., Tremaine, S. 1996, ApJL, 471, L79
	   
\bibitem{}Gebhardt, K., Richstone, D., Kormendy, J., Lauer, T. R.,
Ajhar, E. A., Bender, R., Dressler, A., Faber, S. M., Grillmair, C.,
  Magorrian, J., Tremaine, S. 2000 AJ 119, 1157
   
\bibitem{} Ho L.C., Filippenko A. V., Sargent W. L. W., Peng C. Y. 1997, 
	ApJS 112, 391

\bibitem{}
Irwin J.A  and Bregman J.N 1999 ApJL 510 L21

\bibitem{}
Irwin J.A. and Sarazin C. L.  1998 ApJ 499 650.


\bibitem{} Kim D-W., Fabbiano G., \& Trinchieri G. 1992 ApJ 393, 134.

\bibitem{} Kim D.-W. \& Fabbiano, G. 1995 ApJ 441, 182.

\bibitem{}
Kim D.-W., Fabbiano G., Matsumoto H., Koyama K., Trinchieri G.
1996 ApJ 468, 175

\bibitem{}
Kim D.-W., Fabbiano G., Mackie G. 1998, ApJ. 497, 699.

\bibitem{} Kormendy, J., Bender, R., Richstone, D., Ajhar, E. A.,
Dressler, A., Faber, S. M., Gebhardt, K., Grillmair, C., Lauer, T. R.,
     Tremaine, S. 1996 459, L57

\bibitem{} Manzo, G., Giarrusso, S., Santangelo, A., Ciralli, F.,
 Fazio, G., Piraino, S., Segreto, A. 1997 AAS 122 341

\bibitem{} Matsumoto, H., Koyama, K., Awaki, H., Tsuru, T.,
 Loewenstein, M., Matsushita, K. 1997 ApJ 482 133

\bibitem{} Matsushita, K., Makishima, K., Awaki, H., Canizares, C. R.,
 Fabian, A. C., Fukazawa, Y., Loewenstein, M., Matsumoto, H.,
 Mihara, T., Mushotzky, R. F., Ohashi, T., Ricker, G. R.,
 Serlemitsos, P. J., Tsuru, T., Tsusaka, Y., Yamazaki, T 1994 ApJ 436L 41

\bibitem{} Matsushita, K. et al . (2000) astroph

\bibitem{} Magorrian, J., Tremaine, S., Richstone, D. et al. 1998 AJ 115, 2285

\bibitem{} McElroy D.B. 1995, ApJs 100, 105

\bibitem{}
Oosterbroek T., Parmar A.N., Martin D.D.E., Lammers U. 1997 AA 327 215

\bibitem{} Owens A., Oosterbroek T., Parmar A.N. 1997 AA 324 L9

\bibitem{} Parmar,  A. N., Martin, D. D. E., Bavdaz, M., Favata, F.,
  Kuulkers, E., Vacanti, G., Lammers, U., Peacock, A., Taylor, B. G
  1997 AAS 122 309

\bibitem{} Pellegrini, S. 1994, A\&A 292, 395

\bibitem{} Pellegrini, S. \& Fabbiano, G. 1994, ApJ 429, 105

\bibitem{} Pellegrini, S. 1999, A\&A 343, 23

\bibitem{} Primini F.  et al 1993, ApJ 410 615

\bibitem{} Roberts, T.P.\& Warwick, R.S. 2000, MNRAS 315, 98

\bibitem{} Trinchieri, G., Fabbiano, G. 1991, ApJ 382, 82
310, 637 

\bibitem{} Trinchieri, G., Kim, D-W., Fabbiano, G., \& Canizares, C.R.C. 1994,
ApJ 428, 555

\bibitem{} Trinchieri, G.,  Noris, L., diSerego Alighieri, S. 1997 AA
326, 565

\bibitem{} Trinchieri, G., Israel, G.L., Chiappetti, L., Belloni, T.,
 Stella, L., Primini, F.,
  Fabbiano, P., Pietsch, W. 1999, AA 353, 487

\bibitem{}
van Paradijs J. 1998, ``The many faces of neutron stars", R. Bucchieri,
J. van Paradijs, M.A.Alpar (eds), Kluwer Academic Publishers.



\end{thebibliography}
\end{document}